\documentclass[final,5p,times,twocolumn]{elsarticle}

\usepackage{amssymb}

\journal{Astroparticle Physics}

\usepackage{graphicx}
\usepackage{url}
\include{aas_abbrev}
\usepackage{multirow}

\input{colordvi.tex}

\begin{document}

\begin{frontmatter}



\title{Gamma-Ray Burst Science in the Era of the Cherenkov Telescope Array}

\author[mpik,icrr]{Susumu Inoue\fnref{email}}
\author[open]{Jonathan Granot}
\author[leicester]{Paul T. O'Brien}
\author[tit]{Katsuaki Asano}
\author[ucsc]{Aurelien Bouvier}
\author[inaf]{Alessandro Carosi}
\author[huntsville]{Valerie Connaughton}
\author[iac]{Markus Garczarczyk}
\author[ucsc,sissa]{Rudy Gilmore}
\author[leicester]{Jim Hinton}
\author[kyoto,slac]{Yoshiyuki Inoue}
\author[kek]{Kunihito Ioka}
\author[hiroshima]{Jun Kakuwa}
\author[amsterdam]{Sera Markoff}
\author[icrr,ohio]{Kohta Murase}
\author[leicester]{Julian P. Osborne}
\author[gatech]{A. Nepomuk Otte}
\author[leicester]{Rhaana Starling}
\author[ste]{Hiroyasu Tajima}
\author[icrr,mpp]{Masahiro Teshima}
\author[osaka]{Kenji Toma}
\author[lswhd]{Stefan Wagner}
\author[amsterdam]{Ralph A. M. J. Wijers}
\author[ucsc]{David A. Williams}
\author[konan]{Tokonatsu Yamamoto}
\author[aoyama]{Ryo Yamazaki}
\author{for the CTA Consortium}

\fntext[email]{E-mail: Susumu Inoue (\texttt{sinoue@icrr.u-tokyo.ac.jp}),
 Jonathan Granot (\texttt{granot@openu.ac.il}), Paul O'Brien (\texttt{pto@star.le.ac.uk})}

\address[mpik]{Max-Planck-Institut f\"ur Kernphysik, Saupfercheckweg 1, 69117 Heidelberg, Germany}
\address[icrr]{Institute for Cosmic Ray Research, University of Tokyo, Kashiwanoha 5-1-5, Kashiwa 277-8582, Chiba, Japan}
\address[open]{Department of Natural Sciences, The Open University of Israel, 1 University Road, POB 808, Raanana 43537, Israel}
\address[leicester]{Department of Physics and Astronomy, University of Leicester, University Road, Leicester LE1 7RH, UK}
\address[tit]{Interactive Research Center of Science, Tokyo Institute for Technology, 2-12-1 Ookayama, Meguro-ku, Tokyo 152-8550, Japan}
\address[ucsc]{Santa Cruz Institute for Particle Physics, University of California, Santa Cruz, CA 95064, USA}
\address[inaf]{Istituto Nazionale di Astrofisica, 00136 Roma, Italy}
\address[huntsville]{Department of Physics, University of Alabama in Huntsville, Huntsville, AL 35805, USA}
\address[iac]{Instituto de Astrof\'isica de Canarias, 38200 La Laguna, Tenerife, Spain}
\address[sissa]{Scuola Internazionale Superiore di Studi Avanzati (SISSA), via Bonomea 265, 34136 Trieste, Italy}
\address[kyoto]{Department of Astronomy, Kyoto University, Oiwake-cho, Kitashirakawa, Sakyo-ku, Kyoto 606-8502, Japan}
\address[slac]{SLAC National Accelerator Laboratory, Stanford University, Stanford, CA, USA}
\address[kek]{Theory Center, Institute of Particle and Nuclear Studies, KEK, 1-1 Oho, Tsukuba 305-0801, Japan}
\address[hiroshima]{Department of Physical Science, Hiroshima University, Higashi-hiroshima 739-8526, Japan}
\address[amsterdam]{Astronomical Institute gAnton Pannekoekh, University of Amsterdam, Postbus 94249, 1090 GE Amsterdam, The Netherlands}
\address[ohio]{Center for Cosmology and AstroParticle Physics, Ohio State University, 191 West Woodruff Avenue, Columbus, OH 43210, USA}
\address[gatech]{School of Physics \& Center for Relativistic Astrophysics, Georgia Institute of Technology, 837 State Street NW, Atlanta, GA 30332-0430, USA}
\address[ste]{Solar-Terrestrial Environment Laboratory, Nagoya University, Nagoya 464-8601, Japan}
\address[mpp]{Max-Planck-Institut f\"ur Physik, F\"ohringer Ring 6, 80805 M\"unchen, Germany}
\address[osaka]{Department of Earth and Space Science, Osaka University, Toyonaka 560-0043, Osaka, Japan}
\address[lswhd]{Landessternwarte, Universit\"at Heidelberg, K\"onigstuhl, 69117 Heidelberg, Germany}
\address[konan]{Department of Physics, Konan University, 8-9-1 Okamoto, Higashinada-ku, Kobe, Hyogo 658-8501, Japan}
\address[aoyama]{Department of Physics and Mathematics, Aoyama Gakuin University, Sagamihara 252-5258, Japan}

\begin{abstract}
We outline the science prospects for gamma-ray bursts (GRBs) with the Cherenkov Telescope Array (CTA),
the next-generation ground-based gamma-ray observatory operating at energies above few tens of GeV.
With its low energy threshold, large effective area and rapid slewing capabilities,
CTA will be able to measure the spectra and variability of GRBs at multi-GeV energies with unprecedented photon statistics,
and thereby break new ground in elucidating the physics of GRBs, which is still poorly understood.
Such measurements will also provide crucial diagnostics of ultra-high-energy cosmic ray and neutrino production in GRBs,
advance observational cosmology by probing the high-redshift extragalactic background light
and intergalactic magnetic fields,
and contribute to fundamental physics by testing Lorentz invariance violation with high precision.
Aiming to quantify these goals, we present some simulated observations of GRB spectra and light curves,
together with estimates of their detection rates with CTA.
Although the expected detection rate is modest, of order a few GRBs per year,
hundreds or more high-energy photons per burst may be attainable once they are detected.
We also address various issues related to following up alerts from satellites and other facilities with CTA,
as well as follow-up observations at other wavelengths.
The possibility of discovering and observing GRBs from their onset including short GRBs
during a wide-field survey mode is also briefly discussed.
\end{abstract}

\begin{keyword}
gamma-ray bursts \sep high-energy gamma rays \sep Cherenkov telescopes
\sep cosmic rays \sep cosmology \sep special relativity
\end{keyword}

\end{frontmatter}



\section{Introduction}
\label{sec:intro}

Gamma-Ray Bursts (GRBs) are the most luminous explosions in the Universe after the Big Bang,
liberating as much as $10^{52}-10^{54}$ erg of isotropic-equivalent energy
during a brief period of 0.01-1000 s,
primarily as MeV-band gamma rays.
They are also the most violent explosions,
manifesting rapid and irregular variability on timescales down to sub-millisecond levels.
Since their discovery in 1967, research on GRBs has steadily intensified,
witnessing particularly rapid progress during the last 10-20 years,
driven by observational results from satellite instruments such as
the Burst And Transient Source Experiment (BATSE)
and the Energetic Gamma Ray Experiment Telescope (EGRET)
onboard the {\it Compton Gamma-Ray Observatory (CGRO)},
the {\it High Energy Transient Explorer (HETE-2)}, {\it BeppoSAX}, {\it Swift},
and most recently the Large Area Telescope (LAT) and Gamma-ray Burst Monitor (GBM)
onboard the {\it Fermi} Gamma-ray Space Telescope.
We now know with confidence that:
1. They occur at cosmological distances, typically at redshifts of a few.
2. They are generated by (likely collimated) outflows with ultrarelativisic bulk velocities.
3. Their prompt, MeV-band emission
is accompanied by afterglows that span the radio to X-ray bands
and gradually decay over hours to days or more,
most likely emitted by high-energy electrons accelerated in the blastwave
resulting from the interaction of the outflow with the ambient medium.
4. Those with durations longer than $\sim$2 seconds (``long'' GRBs)
exhibit properties systematically different from those with shorter durations (``short'' GRBs).
5. At least some long GRBs are associated with the core-collapse supernova events of massive stars.
(For recent reviews on GRBs, see e.g. \cite{Piran05,Meszaros06,Woosley06,Gehrels09,Meszaros12}.)

However, many other basic aspects are still unknown or unclear,
such as the identity and nature of the central engine,
the formation mechanism of the ultrarelativistic jet,
the physical mechanisms of energy dissipation and particle acceleration therein
as well as the prompt and early afterglow emission, their cosmological evolution,
the progenitors of short GRBs, etc.
Thus they remain one of the most enigmatic phenomena
in the Universe, and their origin is among the most important unsolved
problems in modern-day astrophysics.

GRBs are also of potentially great importance for other fields of physics and astrophysics.
From model-independent considerations,
they are thought to be one of the leading candidates
for the sources of ultra-high-energy cosmic rays (UHECRs) with energies up to $\sim 10^{20}$ eV,
the highest energy particles known to exist in the Universe today and whose origin is yet unknown \cite{Waxman06}.
The production of UHECRs in GRBs may also induce observable fluxes of high-energy neutrinos.
GRBs are also crucial probes for observational cosmology,
as they are known to occur and are observable out to extremely high redshifts,
including the epochs of cosmic reionization and the earliest star formation \cite{Bromm07}.
Indeed, the recently detected GRB at $z \sim 8.2$ is one of the most distant
and hence most ancient astrophysical objects known to humankind \cite{Tanvir09,Salvaterra09}.
Finally, they can serve as valuable beacons for testing fundamental physics,
particularly in searching for possible violations of Lorentz invariance \cite{Ellis12,Doro12}.

The latest observational advances in GRBs have been brought forth by {\it Fermi} \cite{Granot10}.
The {\it Fermi} LAT instrument has revealed intense emission in the GeV band
from a sizable number of GRBs of both the long and short classes.
The lack of apparent high-energy spectral cutoffs has allowed important new constraints
to be derived on the bulk Lorentz factor of the emitting region.
Some notable, common characteristics have also been discerned
for the GeV emission compared to the MeV emission,
such as the slightly delayed onset, occasionally distinct hard spectral components,
temporally extended emission, etc., for which numerous theoretical explanations have already been proposed.
However, the generally limited statistics of high-energy photons detected by {\it Fermi} LAT
(only a few photons above 10 GeV even in the best cases) have so far
prevented firm conclusions to be drawn on the nature of the high-energy emission from GRBs.

In order to stimulate further progress,
observations with higher sensitivity over a wider energy band are strongly desirable.
Compared to {\it Fermi},
ground-based, imaging atmospheric Cherenkov telescopes (IACTs)
have a large advantage in terms of sensitivity for gamma rays above several tens of GeV
because of their much larger effective area, 
although their field of view and duty cycle are more limited \cite{Aharonian08,Hinton09}.
Past and ongoing follow-up efforts of GRB alerts by current IACTs
such as HESS, MAGIC and VERITAS have yet to uncover signals,
but their present operational threshold energies of $\lesssim$50-100 GeV
and the potential attenuation by the extragalactic background light (EBL) \cite{Krennrich12,Mazin12} in this band
at the typical distances of GRBs could be hindering their detection.

The Cherenkov Telescope Array (CTA), an advanced, next generation ground-based facility
\footnote{In the current timeline, start of construction is foreseen from 2015
and that of scientific operations from 2016 \cite{Hofmann12}.},
is planned to be two sets of mixed arrays of large-size, mid-size and small-size telescopes
(LSTs, MSTs and SSTs, respectively), one each situated in the northern and southern hemispheres
\footnote{The CTA North Array may comprise only the LSTs and MSTs and have somewhat lower
sensitivity above few tens of TeV compared to the CTA South Array.},
which when combined will cover the entire sky over a broad energy range
from tens of GeV up to hundreds of TeV,
with a sensitivity considerably better than existing instruments \cite{CTA11,Hofmann12}.
The most critical component for GRB observations will be the LSTs,
primarily responsible for the lower energy bands.
Compared to current IACTs, they will feature:
i) appreciably lower threshold energy ($\lesssim$30 GeV, possibly down to $\sim$ 15 GeV in some cases),
and ii) even larger effective area at multi-GeV energies ($\sim 10^4$ times larger than {\it Fermi} LAT at 30 GeV) \cite{Funk12}.
In addition, they are designed with iii) rapid slewing capability (180 degrees azimuthal rotation in 20 seconds),
comparable to MAGIC-II, allowing the observation of some long GRBs during their prompt phase,
and many others in the early afterglow phase.
By acquiring high-quality (i.e. high photon statistics) measurements of
time-resolved spectra and energy-dependent variability at multi-GeV energies
that was not possible with {\it Fermi}, some important science goals
that can be addressed with CTA include the following.
1. Determine or more robustly constrain the bulk Lorentz factor of the emission zone.
2. Determine the emission mechanisms of prompt GRBs and early afterglows.
3. Reveal hadronic signatures accompanying the production of UHECRs and neutrinos
4. Probe the extragalactic background light
at high redshifts, beyond the reach of blazar active galactic nuclei ($z \gtrsim 2$).
5. Probe Lorentz invariance violation with better precision.

This article aims to provide an overview of the science prospects for GRBs with CTA,
and is organized as follows.
We begin by reviewing our current knowledge of GRBs, focusing on their emission
in the high-energy ($\gtrsim$ 100 MeV) and very-high-energy ($\gtrsim$ 100 GeV) gamma-ray regimes
in Sections \ref{sec:status-GeV} and \ref{sec:status-VHE}, respectively.
In Section \ref{sec:science-case}, selected science cases for CTA are described in some detail.
Section \ref{sec:simulation}  presents demonstrative simulations of GRB spectra and light curve measurements,
as a first step toward quantitative assessments of the science goals.
In Section \ref{sec:rates}, predictions for GRB detection rates are given from two different perspectives.
Section \ref{sec:followup} discusses various issues related to following up GRB alerts with CTA and at other wavelengths,
as well as the possibility of discovering GRBs with CTA alone during a wide-field survey mode.
We conclude and provide an outlook in Section \ref{sec:conc}.


\section{Current Status at GeV Energies}
\label{sec:status-GeV}

\subsection{From EGRET to Fermi}
\label{sec:pre-Fermi}

GeV emission from GRBs was first discovered by EGRET on-board {\it CGRO}, active during 1991$-$2000.
While EGRET detected only five GRBs with its spark chambers within 20$\,$MeV$-$30$\,$GeV
and a few more bursts with its Total Absorption Shower Counter within $1-200\;$MeV,
these events already showed diversity \cite{Dingus01}.
For GRB~940217, GeV emission was seen up to $\sim 1.5\;$hr after the burst trigger,
including an $18\;$GeV photon at $\sim 1.3\;$hr~\cite{Hurley94}.
GRB~941017 displayed a distinct, high-energy spectral component
up to $\sim 200\;$MeV with a hard spectrum of $\nu F_\nu \propto \nu$~\cite{Gonzalez03},
long duration $\sim 200\;$s and total energy $\sim 3$ times larger
than the low-energy, MeV-band spectral component that lasted several tens of seconds.
A promising explanation is inverse-Compton emission from the forward-reverse shock system
that forms as the ultra-relativistic GRB outflow is decelerated by the external medium~\cite{GrGu03,PW04}.
However, better data are needed in order to conclusively determine the origin of such high-energy components.
The Italian experiment Astro-rivelatore Gamma a Immagini LEggero ({\it AGILE}), launched in 2007,
detected emission from GRB~080514B up to $\sim 300\;$MeV
that continued for $>\,$13$\;$s, compared to $\sim 7\;$s for the $\sim\,$MeV component ~\cite{Giuliani08}.  

Significant progress was brought forth by the {\it Fermi} Gamma-ray Space Telescope,
which has been detecting about $\sim 8\,$--$\,10$ GRBs per year at energies above a few tens of MeV
since its launch on 11 June 2008.
Thanks to the very wide energy range covered by its instruments,
GBM (8$\;$keV - 40$\;$MeV)
and LAT (25$\;$MeV - $>\,$300$\;$GeV),
the prompt emission spectra of some GRBs have been measured
from $\sim 8\;$keV up to $\sim 30\;$GeV.
Below is a brief summary of what it has taught us about 
GRBs and associated high-energy physical processes.

\subsection{Energetics \& average spectrum}
\label{sec:energetics+spectrum}

From a comparison of the number of GRBs per year detected by {\it Fermi} LAT
with expectations based on simple spectral extrapolation of BATSE results \cite{Band09,Granot10},
as well as the observed flux or upper limits in the GeV-band
relative to the MeV flux for individual bursts \cite{GPW11,Beniamini11,Ackermann12},
the energy output in the GeV band appears to be about an order of magnitude less than in the MeV band.
In some cases, a high-energy extrapolation of the Band function spectrum \cite{Band93}
overpredicts the GeV flux, indicating the presence of a spectral steepening or cutoff.
This is despite the fact that the spectra of many bright GRBs are well described with the
Band function over the entire {\it Fermi} energy range \cite{Zhang:2010ey},
and the brightest LAT GRBs often exhibit a hard, distinct high-energy spectral component
(Section \ref{sec:delay+HEcomp}).
At any rate, this suggests that in most GRBs
the radiative output at GeV energies
s not a major fraction of the total energy budget.
However, short GRBs may be different in this respect (Section \ref{sec:long-vs-short}).

\subsection{Constraints on the Lorentz factor}
\label{sec:bulkG}

In addition to its typically enormous isotropic-equivalent luminosities $L \sim 10^{50}-10^{53}\;{\rm erg\;s^{-1}}$,
the prompt emission of GRBs
shows significant short timescale variability,
with a good fraction of the radiated energy in photons with energies $E_{\rm ph} \gtrsim m_e c^2$.
These properties would imply a huge optical depth to pair production
($\gamma\gamma\to e^+e^-$) at the source, which would thermalize the
spectrum and be at odds with the observed non-thermal spectrum,
unless the emitting material was moving toward us ultrarelativistically
with a bulk Lorentz factor $\Gamma\gg 1$.
Such ``compactness'' arguments had been applied to EGRET GRBs
to derive lower limits
on the value of $\Gamma$, typically resulting in $\Gamma_{\rm min} \sim 10^2$,
and in some cases as high as a few hundred (see~\cite{LS01} and references therein).
However, these limits for EGRET bursts were based on the implicit assumption that the spectra
extended to well above the observed energy range,
even though direct evidence was lacking for photons with energies high enough
that substantiate the opacity constraint.

For {\it Fermi} LAT GRBs, one could adopt a more robust approach
of relying only on photons within the directly observed energy range.
If $E_{\rm ph,max}$ is the highest observed photon energy, 
the maximal value of $\Gamma_{\rm min}$  that can be derived
corresponds to the case in which this energy is just above the pair production threshold
in the comoving frame of the flow where the photons are roughly isotropic, so that
\begin{equation}\label{eq:bulkG}
\Gamma_{\rm min} \lesssim (1+z)\frac{E_{\rm ph,max}}{m_ec^2} 
\approx 2000 (1+z)\left(\frac{E_{\rm ph,max}}{1 \,{\rm GeV}}\right)\ .
\end{equation}
Provided that photons of sufficiently high energies are detected,
concrete values for $\Gamma_{\rm min}$ can be obtained
by relating the spatial scale of the emitting region to the observed timescale of variability $t_v$,
the exact choice of which constitutes the main uncertainty for $\Gamma_{\rm min}$.
Other uncertainties arise from those on the spectral fit parameters,
or on the degree of space-time overlap between the high-energy photon
and lower energy photons relevant for pair production.
Relying on a single high-energy photon would also induce an uncertainty
as it can still escape from regions with an optical depth of up to a few,
but accounting for the second or third highest-energy photons helps to mitigate such uncertainties
due to Poisson fluctuations.
Thus, reasonably robust values have been derived for 3 of the brightest LAT GRBs:
$\Gamma_{\rm min} \approx 900$ for GRB~080916C~\cite{080916C},
$\Gamma_{\rm min} \approx 1200$ for GRB~090510~\cite{090510-phys},
and $\Gamma_{\rm min} \approx 1000$ for GRB~090902B~\cite{090902B}.  

Nevertheless, these limits rely on the simple assumption
of a photon field that is homogeneous, time-independent
and isotropic in the rest frame of the emitting material.
A fully time-dependent and self-consistent calculation
featuring emission from a thin spherical shell over a finite range of radii~\cite{Granot08},
appropriate for the internal shock model,
gives limits that are lower by a factor of $\sim 3$.
Similar conclusions were recently reached by
several studies \cite{Li:2008ub,Aoi:2009ty,Zou11,HDMV12}
(see also Section \ref{sec:GRBphys}.)

In one case so far, GRB~090926A~\cite{090926A},
a high-energy break or cutoff at $E_{\rm ph,cut} \sim 1.4\;$GeV 
was measured at the brightest part of the prompt emission.
If this is due to internal pair opacity at the source,
for the first time we can determine the Lorentz factor
instead of just setting a lower limit,
We arrive at $\Gamma \approx 720\pm 76$ for a simple one-zone model,
and $\Gamma \approx 220$ for the fully time-dependent model mentioned above~\cite{Granot08}.
However, the available statistics of high-energy photons is insufficient
to ascertain whether the measured spectral softening is actually caused by pair opacity
or instead has a different physical origin.

The fact that $\Gamma_{\rm min}$ for the bright, short GRB~090510 is comparable
and even slightly higher than those for the bright, long GRBs 080916C and 090902B
demonstrates that short GRBs are as ultrarelativistic as long GRBs,
which was questioned before the launch of {\it Fermi} \cite{Nakar07}.
Note also that since the highest values of $\Gamma_{\rm min}$
are derived for the brightest LAT GRBs, they may be affected by selection effects.
For example, GRBs with higher $\Gamma$ may tend to be brighter in the LAT band
by avoiding internal pair production~\cite{Ghisellini10}.

\subsection{Delayed onset \& distinct high-energy spectral component}
\label{sec:delay+HEcomp}

A common feature in {\it Fermi} LAT GRBs is a delay in the onset of the
emission $>\,$100$\;$MeV relative to the emission $\lesssim 1\;$MeV.
Such a delayed onset clearly appears in the four brightest LAT bursts so far,
while in dimmer LAT bursts it is often inconclusive because of poor photon statistics.
The time delay $t_{\rm delay}$ appears to scale with the duration of the GRB;
$t_{\rm delay}\sim$ several seconds in the long GRBs 080916C and 090902B,
while $t_{\rm delay} \sim 0.1-0.2\;$s in the short GRBs 090510 and 081024B
\cite{090510-phys,081024B}, albeit with a smaller statistical significance for the latter.

Only 3 LAT GRBs so far have shown clear ($>\,$5$\,\sigma$) evidence
for a distinct, high-energy spectral component.
However, these GRBs are the 3 brightest in the LAT,
while the next brightest, GRB 080916C, showed a hint for an excess at high energies.
This suggests that such distinct high-energy spectral components may be common,
but can only be detected with high significance in particularly bright cases.
The distinct spectral component is usually well fit by a hard power-law dominating at high energies.
In GRB~090902B, a single power-law component dominates over the usual Band component
not only above $\sim 100\;$MeV, but also below $\sim 50\;$KeV.
There is also marginal evidence for a similar low-energy excess in GRB~090510.

Both the delayed onset and distinct spectral component should be
strongly related to the prompt GRB emission and may help elucidate its uncertain radiation mechanism.
We discuss here a few selected models, of both leptonic and hadronic types.
Further alternative models are described in Sections \ref{sec:GRBphys} and \ref{sec:UHECR}.

\noindent
{\bf Leptonic.}
One might attribute the high-energy spectral component to inverse-Compton emission,
in particular the synchrotron-self Compton (SSC) mechanism,
if the main MeV-band component is of synchrotron origin
~\cite{Sari96,Papathanassiou96,Pilla98,Bosnjak09,asa11}.
However, it may be difficult to account for the observed $t_{\rm delay} > t_v$,
the width of individual spikes in the light curve.
Moreover, the observed, gradual steepening with time
of the high-energy spectral component is not naturally expected.
The fact that the spectral indices of the distinct GeV component and the sub-MeV component
are notably different, as well as the existence of the low-energy excess components in some bursts,
are also problematic for the simplest leptonic models.

\noindent
{\bf Hadronic.}
One might identify $t_{\rm delay}$ with the acceleration time $t_{\rm acc}$
of protons (or heavier ions) up to ultra high energies,
where they can convert their energy to GeV photons with a reasonable efficiency,
e.g. via the proton synchrotron mechanism~\cite{Razzaque10}.
If the GeV emission involves photohadronic cascades (i.e. inverse-Compton emission
by secondary $e^\pm$ pair cascades initiated by interactions between ultra-high-energy
protons and low-energy photons ~\cite{Asano09}), some additional delay may occur
for the cascades to develop.
However, such a scenario for the delayed onset requires the GeV emission
to originate from the same physical region over times $> t_{\rm delay}$
and implies variability timescales $t_v \gtrsim t_{\rm acc} \sim t_{\rm delay}$
due to the stochastic nature of the acceleration process,
which is in contrast to what is usually observed, $t_v < t_{\rm delay}$.
The gradual steepening of the high-energy spectral component is evidently not expected,
although it might be mimicked by time evolution effects~\cite{Razzaque10}.
For GRB~090510, a photohadronic cascade model requires a total isotropic equivalent energy
$> 10^2$ times larger than in MeV gamma rays~\cite{Asano09},
potentially posing a challenge for models of its progenitor.
On the other hand, the low-energy excess component observed in GRB~090902B (and possibly also in GRB~090510)
may be naturally explained in this model by synchrotron emission from secondary pairs~\cite{Asano09,Asano10}.

\subsection{Long-lived high-energy emission}
\label{sec:long-lived}

The GeV-band emission in most {\it Fermi} LAT GRBs
lasts significantly longer than the MeV-band emission.
During the prompt phase of the MeV emission,
the GeV emission usually shows significant variability,
in some cases being correlated with the former.
The longer-lived GeV emission, however, is temporally smooth
and consistent with a power-law flux decay,
typically $\propto t^{-1.2}-t^{-1.5}$ but ranging over $\sim t^{-0.8}-t^{-1.9}$,
with roughly flat $\nu F_\nu$ spectra.

The early GeV emission may be most naturally interpreted as
the high-energy counterpart of the prompt MeV emission from the same emission region, 
especially when the respective light curves show temporal correlation.
The long-lived, smooth decay phase is more naturally ascribed
to the high-energy component of the afterglow emission from the external forward shock.
(For discussion of late-time flaring emission, see Section \ref{sec:GRBphys}.)
Although an afterglow origin has been suggested in some cases for the entire GeV emission including the prompt phase
~\cite{KB-D09,Ghisellini10},
it is generally hard to explain the sharp spikes seen in the early LAT light curve \cite{He+11}.
Good multi-wavelength coverage of the afterglow from early times (e.g. ~\cite{090510-AG})
should be a key to identify its origin.

Producing multi-GeV photons is challenging for a leptonic synchrotron origin,
both during the prompt emission~\cite{080916C},
and even more so during the afterglow (e.g.~\cite{LW06}; see however, \cite{KB-D09})
as it requires a very high bulk Lorentz factor and high upstream magnetic field,
in addition to very efficient shock acceleration of electrons.
For example, the photon with energy $33\;$GeV ($94\;$GeV in the cosmological rest frame at redshift $z=1.822$)
observed in GRB~090902B at $82\;$s after the trigger,
well after the end of the prompt emission~\cite{090902B}, would require $\Gamma > 1500$ in this interpretation.
That at least some GRBs are now known to emit photons near 100 GeV
is also a crucial fact for ground-based gamma-ray observatories (Section \ref{sec:status-VHE}).
See Section \ref{sec:GRBphys} for further discussions on the high-energy afterglow emission.

\subsection{High-energy emission of long versus short GRBs}
\label{sec:long-vs-short}

Out of the 30 LAT GRBs up to January 2011,
4 and 26 are of the short and long duration class, respectively.
Bearing in mind the uncertainty due to the small number statistics,
this implies that $\sim 13\%$ of LAT GRBs are short,
which is marginally consistent with the $\sim 20\%$ fraction of short GRBs detected by the GBM.
The GeV emission properties of short and long GRBs appear to be rather similar.
Both can produce very bright emission in the LAT energy range
(\cite{090510-phys} vs. \cite{080916C,090902B,090926A}),
with a correspondingly high lower limit on the bulk Lorentz factor,
$\Gamma_{\rm min} \sim 10^3$ for a simple one-zone model
or $\sim 10^{2.5}$ for a time-dependent model relevant for internal shocks,
as well as a distinct spectral component (\cite{090510-phys} vs. \cite{090902B,090926A}).
Both show delayed onset and long-lived GeV emission compared to the MeV emission
(see also \cite{Giuliani10} for {\it AGILE} results).
However, the GeV onset delay time appears to roughly scale with the duration of the GRB,
being $\sim 0.1-0.2\;$s for short GRBs and several seconds for long GRBs.
Considering that GRB 080916C and GRB 090510 had comparable isotropic equivalent luminosities
of several times $10^{53}\;{\rm erg\,\,s^{-1}}$,
this scaling of the delay times may suggest an underlying cause
that is largely independent of the outflow's instantaneous properties
and is instead driven by its general evolution over the duration of the GRB,
which differs between short and long GRBs and
may provide clues to the inner workings of their central engines,
e.g.~\cite{Toma11}.
Another interesting potential difference is that the GeV energy output relative to the MeV output
of short GRBs appear to be comparable, while that of long GRBs tend to be a smaller fraction.
However, this still warrants confirmation,
since there are only 4 short LAT GRBs so far that may be subject to selection effects.


\section{Current Status at Very High Energy}
\label{sec:status-VHE}

Because of their limited effective area, the sensitivity of satellite instruments
is often inadequate to measure the decreasing fluxes from gamma-ray sources
above few tens of GeV.
In this very high energy (VHE) regime, 
ground-based Imaging Atmospheric Cherenkov Telescopes (IACTs)
are the most sensitive instruments.
GRB follow-up observations are regularly carried out with
the latest generation of IACTs including
the Major Atmospheric Imaging Cherenkov Telescope (MAGIC)\footnote{\url{http://wwwmagic.mppmu.mpg.de}},
the High Energy Stereoscopic System (H.E.S.S.)\footnote{\url{http://www.mpi-hd.mpg.de/hfm/HESS/}}, and
the Very Energetic Radiation Imaging Telescope Array System (VERITAS)\footnote{\url{http://veritas.sao.arizona.edu/}}.
Furthermore, facilities such as Milagro\footnote{\url{http://www.lanl.gov/milagro/}}
and ARGO\footnote{\url{http://argo.na.infn.it/}}
have been conducting searches for VHE emission from GRBs
that are complementary to IACTs,
with much wider field of view and higher duty cycle,
albeit with less sensitivity and higher energy threshold
(see e.g. \cite{Atkins05,Abdo07,Aune09,Aielli09a,Aielli09b} for selected results).

\subsection{MAGIC}
\label{sec:MAGIC}

In view of the expected attenuation of gamma rays due to the EBL (\cite{Krennrich12,Mazin12}, see also Section \ref{sec:EBL})
and the generally large redshifts of GRBs,
achieving the lowest possible energy threshold for ground based detectors is of paramount importance.
Furthermore, since the duration of the prompt emission is typically tens of seconds,
fast repositioning of the telescope towards the coordinates provided by satellite detectors is crucial.
Both requirements are met with the two MAGIC telescopes. 
With reflectors of 236~m$^2$ each, they are currently the largest stereoscopic pair of IACTs in the world,
aiming to explore the gamma-ray sky with high sensitivity at energies starting well below 100~GeV.
Moreover, the lightweight design of their supporting cradle allows MAGIC to slew
$180^\circ$ in azimuth in less than 20~s.

Having a limited field of view, IACTs must generally rely on external GRB alerts
such as those provided by the automated satellite link to the Gamma-ray Burst Coordinate Network
(GCN)\footnote{\url{http://gcn.gsfc.nasa.gov/}},
which broadcasts the coordinates triggered by selected satellite detectors and sends them to ground based experiments.
The GCN information can be received directly over a TCP/IP internet socket connection.
A program is necessary to validate the alert with predefined observability criteria.
In the case of MAGIC, the alert is automatically accepted and sent directly to the central control software
if the following criteria are fulfilled:

\begin{itemize}
\item The Sun is below the astronomical horizon (zenith $> 103^\circ$).
\item The angular distance from the GRB to the Moon is $> 30^\circ$.
\item The zenith angle for the GRB observation is $< 60^\circ$.
Under moonlight the maximal zenith angle is reduced to $55^\circ$.
\end{itemize} 

\noindent
Because of their large localization uncertainties, {\it Fermi} GBM alerts are not followed up by many ground based telescopes.
In order to increase the chances for simultaneous observations with MAGIC and {\it Fermi} LAT,
some GBM alerts are accepted according to the following criteria:

\begin{itemize}
\item Flight generated: error $< 4^\circ$, signal-to-noise $> 100$, hardness ratio (counts at 15-50keV relative to 50-300keV) $< 1$
\item Ground generated: error $< 4^\circ$, signal-to-noise $> 40$.
\item The pointing is updated if more precise coordinates arrive.
\item Abort of the observation after 1~h if error $> 1.5^\circ$.
\end{itemize}

\begin{figure}[!htb]
\centerline{\hspace{0.9cm}\includegraphics[width=1.1\columnwidth]{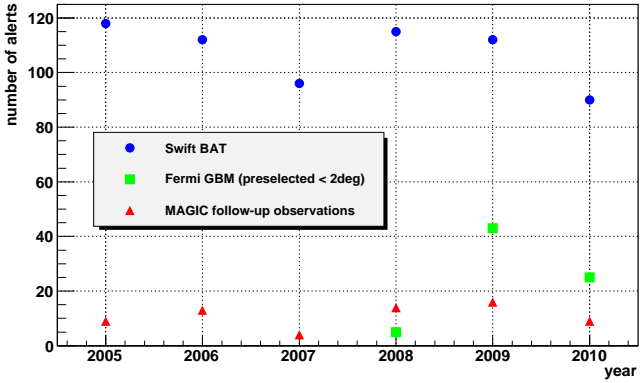}}
\caption{Frequency of alerts distributed by GCN and those followed up by MAGIC during 2005 - 2010.}
\label{fig:alertfrequency}
\end{figure}

Nominal duration of observations is from the start of the observability until 4 hours after burst trigger, $T_{0}+4 \mbox{h}$.
As the redshift of the source is only known hours to days later, one is obliged to observe all candidates,
although the usefulness of the observation can be limited if later follow-up reveals the redshift to be too high,
or if it turns out that the redshift could not be measured.
Figure~\ref{fig:alertfrequency} shows the frequency of alerts received by MAGIC in the time period between 2005 and 2010.
On average 105.4 alerts from BAT and GBM (with some additional preselection) are received per year.
Out of this sample, on average 9.6 follow-up observations are performed, which implies a duty cycle of 11\%.

MAGIC has followed up 70 GRB alerts up to June 2012.
In 23 cases the redshift of the burst was measured and out of this sample 5 GRBs had $z < 1$.
Unfortunately, because of delays caused by observability criteria or technical problems,
most of the MAGIC data sets were obtained only well after the GRB prompt emission phase.
So far no VHE emission components have been detected.
Some early MAGIC results on GRBs were presented in~\cite{Albert07}.

Two MAGIC follow-up observations can be highlighted here,
conducted with the single, MAGIC-I telescope.
For GRB 080430~\cite{grb080430magic} and GRB 090102~\cite{grb090102magic},
simultaneous multiwavelength data and the knowledge of the redshift have allowed
detailed discussions on the physical implication of the VHE upper limits.

\begin{figure}[!htb]
\centerline{\hspace{0.9cm}\includegraphics[width=1\columnwidth]{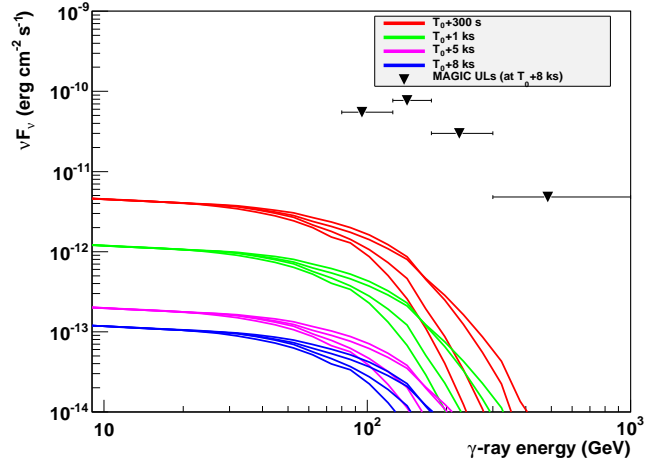}}
\caption{
95\% confidence level ULs derived by MAGIC at various energies as black triangles,
compared with predictions of the SSC emission for the afterglow of GRB 080430 at different time delays after the burst onset.
Curves with the same color refer to the same emission model, but with different models for attenuation by the EBL.
The time delay for the blue curves correspond to the MAGIC observation window. From~\cite{grb080430magic}.}
\label{fig:grb080430}
\end{figure}

GRB 080430 occurred while the Sun was still above the horizon at the MAGIC site.
The MAGIC observation started at $T_{0} + 4753$~s, well after the end of the prompt emission phase,
but under very good observing conditions.
Analysis of the data set resulted in upper limits (ULs) starting at 80~GeV.
Figure~\ref{fig:grb080430} shows these MAGIC ULs in comparison
with model expectations of the afterglow emission for this event.

\begin{figure}[!htb]
\centerline{\hspace{0.9cm}\includegraphics[width=0.95\columnwidth,angle=0]{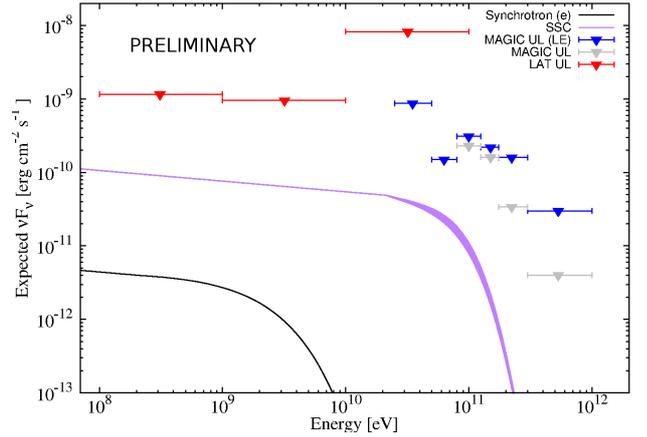}}
\caption{Simultaneous, 95\% CL upper limits derived by MAGIC as blue triangles and by {\it Fermi} LAT as red triangles,
compared with the expected SSC emission for the afterglow of GRB 090102.
The shaded region denotes the uncertainty in EBL attenuation. From~\cite{grb090102magic}.}
\label{fig:grb090102}
\end{figure}

The follow-up observation of GRB 090102 by MAGIC started at $T_{0}+1161$~s, while {\it Swift} and {\it Fermi}
were still on the target.
For the first time, simultaneous ULs with LAT and MAGIC could be extracted (Figure~\ref{fig:grb090102}),
showing the potential impact of ground-based VHE observations that cover the energy range
where EBL attenuation is expected to be substantial.

\subsection{H.E.S.S.}
\label{sec:HESS}

The H.E.S.S. I array consists of four telescopes with a mirror collection area of 108m$^2$ each,
operational in Namibia since 2004.
Their energy threshold depends on the zenith distance
of the observing position and reaches 100 GeV at zenith.
In 2012, the array was complemented with the single H.E.S.S. II telescope with a 600m$^2$ primary mirror,
the largest Cherenkov telescope ever built,
aiming for a significantly lower energy threshold.
The discussion below concerns observations conducted to date with the smaller, H.E.S.S. I telescopes.

While small compared to the LAT instrument for lower energy gamma-rays,
the field-of-view of the H.E.S.S. telescopes has a diameter of 5 degrees,
significantly larger than those of other past and current IACTs.
This is particularly beneficial in the context of GRB observations.
Several hard-X-ray satellite-borne GRB instruments provide
coarse GRB location accuracies at the first trigger signal.
Subsequent on-board or ground analysis improves the positional accuracy with significant latency.
All but one GRB trigger followed up by the H.E.S.S. telescopes had error circles associated
with the initial trigger which were fully covered by the field-of-view of the H.E.S.S. telescopes.
The H.E.S.S. GRB programme follows up on triggers promptly,
whenever the location can be observed with the H.E.S.S. array.
This requires the GRB trigger to arrive when the telescopes are operational
(Sun at least 15 degrees below the horizon, no moonlight, clear skies),
and a burst location at trigger corresponding to a zenith angle less than 45 degrees.
The duration of the observations extends for at least one hour and depends further
on the position of the GRB trigger, potential signals in on-line analysis,
and estimates of the GRB redshift as reported on GCN alerts or other sources during ongoing H.E.S.S. observations.
Triggers from any instrument feeding the GCN trigger system are accepted.
In addition a GRB afterglow observing mode is followed,
observing GRB locations as they become observable for H.E.S.S. with a latency
of up to 24h after the trigger in case of GRBs with established redshifts.
The latency depends on the GRB redshift and fluence.
Optical follow-up observations of GRBs visible for H.E.S.S. are obtained
with the ROTSE 3c telescope operating at the H.E.S.S. site.

Triggers for prompt observations are passed on directly from
the local socket connection to the telescope operation system.
Since 2010 they interrupt any ongoing observations if observability criteria are met automatically.
This results in a reduction of the latency of the VHE observations.
At any time a GRB shift expert is on-duty, supporting the observers with follow-up investigations
of GRB properties, redshift estimates, and further considerations affecting the GRB observations.
Results of these studies have been published in \cite{Tam1,Tam2,Lennarz1}.
Neither individual GRB observations nor stacked analyses provided the detection of VHE signals.
In the case of GRB 070621 the sensitivity of H.E.S.S. allowed establishing upper limits
on the VHE energy flux (in ${\rm erg \ cm^{-2} \ s^{-1}}$, above 200 GeV) which are
{\it lower} than the {\it detected} X-ray energy flux (0.3 - 10 keV, observed with XRT)
during the period 300 - 3000 sec after the trigger \cite{Tam1}.
Other, particularly noteworthy events are GRB 060602B \cite{Tam2} and GRB 100621A \cite{Lennarz1}.
Observations of GRB 060602B were remarkable because
the trigger occurred within the field-of-view of H.E.S.S. at the time of the trigger -
even if the position was at the edge of the field-of-view where the sensitivity
is significantly reduced compared to on-axis performance.
It was the only event witnessed by an IACT with data being taken {\it before, during, and after} the trigger.
Only upper limits have been obtained \cite{Tam2}.
The unusually soft X-ray spectrum and the low Galactic latitude of the event suggest,
however, that GRB 060602B was actually a hard Galactic X-ray transient.
This notion has been confirmed by \cite{Barthelmy07}.


\subsection{VERITAS}
\label{sec:VERITAS}

Located at the Fred Lawrence Whipple Observatory in southern Arizona, USA,
VERITAS is an array of four IACTs using a Davies-Cotton design with a 12 m dish
and a $3.5^{\circ}$ field-of-view camera composed of 499 photomultiplier tubes.
The VERITAS observing strategy assigns highest priority to GRB observations.

GCN alerts are received at the VERITAS basecamp through a TCP/IP socket connection
which allows prompt notification to the observers.
If the burst localization is higher than $20^{\circ}$ elevation and has less than $10^{\circ}$ localization uncertainty,
the telescopes are immediately slewed toward the least uncertain burst localization.
The telescope slewing rate is $\sim 1^{\circ}$/s both in azimuth and elevation.
Since VERITAS first light in 2006, 36 burst alerts were promptly responded to.
For half of them, data taking could start less than 180 sec after the burst trigger with a record of 91 sec.

VERITAS policy is then to continue observation as long as the target remains above $20^{\circ}$ elevation
and for at least 3 hours in the case of {\it Swift} alerts and 1 hour in the case of {\it Fermi} GBM alerts
(since in most cases, GBM localization uncertainty is much larger than the VERITAS field-of-view).
Additionally, in order to maximize VERITAS chances of detecting a GBM burst,
an observation mode where the GBM 1-$\sigma$ error box is being continuously scanned
is in the process of being implemented \cite{OrbitMode}.

\begin{figure}[!h]
\includegraphics[width=1\columnwidth]{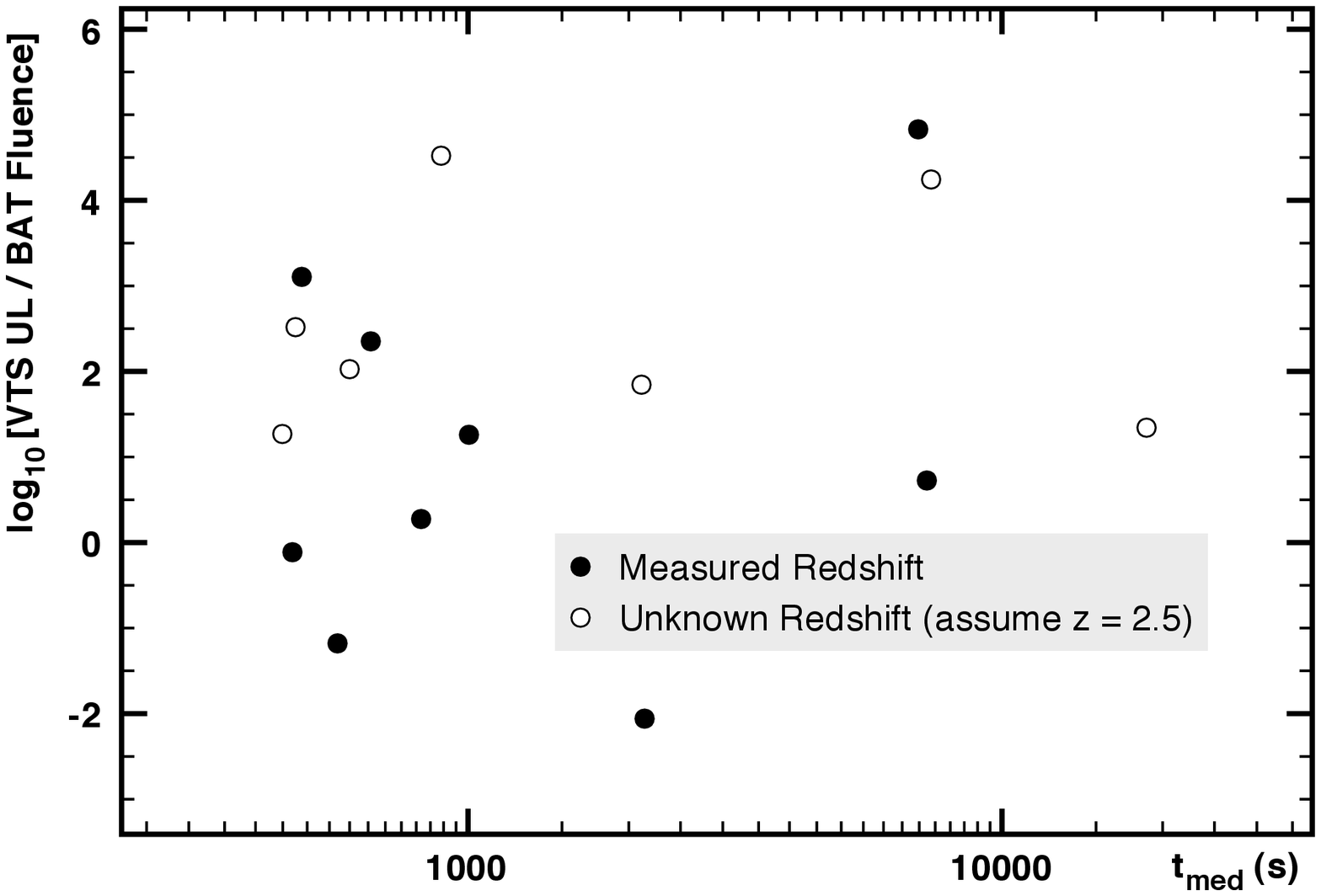}
\caption{EBL-corrected VERITAS integral fluence upper limits above 200 GeV,
divided by the fluence measured by the {\it Swift} BAT in the 15 - 350 keV energy band as a function of $t_{med}$,
the time since the beginning of VERITAS observations at which we expect to detect half of the photon signal
assuming $t^{-1.5}$ temporal decay \cite{GRBSwift}.}
\label{fig:VTSUL}
\end{figure}

Upper-limits for a sample of 16 {\it Swift}-triggered GRBs were recently published \cite{GRBSwift}.
Among the 9 bursts with measured redshifts, 3 could be constrained to have VHE afterglows
less energetic than the prompt, low-energy gamma-rays measured by {\it Swift} in the 15-350 keV range,
as shown in in Fig. \ref{fig:VTSUL}.

\section{GRB Science Cases for CTA}
\label{sec:science-case}

\subsection{Physics of GRBs}
\label{sec:GRBphys}

Many fundamental problems remain unsolved concerning the physical mechanisms behind GRBs.
With its large effective area, CTA can detect hundreds or more photons
from moderate to bright GRBs (Sections \ref{sec:simulation} and \ref{sec:rates})
and achieve unprecedented temporal and spectral resolution
in the domain above a few tens of GeV.
Here we discuss the prospects for studying the physics of GRBs with CTA,
focusing on issues related to the bulk Lorentz factor, the prompt emission, and the afterglow emission.

\vspace{0.3cm}
\noindent{\bf Bulk Lorentz Factor}.
The bulk Lorentz factor $\Gamma$ is a key quantity characterizing ultra-relativistic outflows in GRBs,
particularly their acceleration mechanism and composition
(or degree of baryon loading and magnetization).
The baryon loading is intimately related to jet formation at the central engine,
the most enigmatic aspect of GRB physics.
In the canonical fireball model of GRB outflows, an optically thick plasma of
$e^\pm$ pairs, photons and some baryons expands and accelerates
because of its own pressure and (with sufficient baryon loading)
converts most of the initial internal energy to the kinetic energy of baryons,
a fraction of which is later dissipated through internal shocks \cite{Piran05,Meszaros06}.
The electrons associated with the baryons are essential for trapping the radiation,
and the corresponding minimum amount of baryons implies an upper limit on $\Gamma$ of
\begin{equation}\label{eq:bulkGlimit}
\hspace{-0.5cm}
\Gamma < \left(\frac{L \sigma_T}{4\pi m_p c^3 r_0}\right)^{1/4}
\sim 10^3
\left(\frac{L}{10^{53}\ {\rm erg}\ {\rm s}^{-1}}\right)^{1/4}
\left(\frac{r_0}{10^{7}\ {\rm cm}}\right)^{-1/4},
\end{equation}
where $L$ is the isotropic-equivalent luminosity and $r_0$ is the initial size of 
the fireball \cite{Meszaros:1999gb}.

As discussed in Section \ref{sec:bulkG}, $\Gamma$
can be constrained by identifying the high-energy spectral cutoff
due to internal pair production ($\gamma \gamma \to e^+e^-$).
Probing the range of $\Gamma$ in Eqs.\ref{eq:bulkG} and \ref{eq:bulkGlimit}
will test the fireball model as well as alternatives involving magnetic acceleration
\citep[e.g.,][]{Thompson:1994,Meszaros:1996ww,Spruit:2000zm,
Giannios:2007yj,Komissarov:2008ic,Tchekhovskoy:2009mf,GKS11}.
Values of $\Gamma \gtrsim 1000$ would challenge
both the fireball model and magnetic acceleration models
(in which the acceleration proceeds more slowly with radius)
and point to other possibilities, such as converging jets \cite{Ioka11}
or baryon entrainment during acceleration \cite{Ioka10}.

Although the instantaneous emission from a thin shell would exhibit
a photon spectrum with an exponential cutoff,
$\nu F_{\nu} \propto E^{\lambda} \exp[-\tau_{\gamma\gamma}(E)]$,
where $\tau_{\gamma\gamma}(E)$ is the opacity to pair production
for photons of energy $E$, the shape of the time-integrated spectrum of a single pulse (or multiple pulses)
may depend on the details of the emission mechanism and geometry.
For example, the simple model of an emitting slab would give
\begin{equation}
\nu F_{\nu} \propto 
E^{\lambda} \frac{1-\exp(-\tau_{\gamma\gamma}(E))}{\tau_{\gamma\gamma}(E)},
\end{equation}
a smoothly broken power-law spectrum \cite{Bar06}.
However, this does not account for the realistic contribution to the opacity from 
the radiation field extending outside the emitting region.
It is most likely that time-dependent and/or multi-zone effects
(e.g., the superposition of emission from multiple internal shocks)
modify the spectral break feature \cite{Granot08,Li:2008ub,Aoi:2009ty,HDMV12}.
As for multi-zone effects, the external inverse-Compton (EIC) emission
from internal shock electrons at outer radii upscattering
seed photons from inner radii can be potentially important
and complicate the resulting spectra \cite{Li:2008ub,Zou11,Toma11}.
The softening of the spectral index could become so mild
that observational distinction would be difficult with {\it Fermi},
yet feasible with CTA with much higher photon statistics.

Time-resolved spectral analysis with CTA
can clarify the evolution of $\tau_{\gamma\gamma}$
and the corresponding cutoff or break energy $E_1$ where $\tau_{\gamma\gamma}(E_1) \equiv 1$,
possibly even within a single pulse of the light curve,
as predicted in time-dependent models \cite{Granot08}.
Such analysis is also crucial to distinguish the prompt emission from the afterglow.
The early afterglow emission could possibly dominate over the prompt emission at $E > E_1$
and hinder the determination of the bulk Lorentz factor.
This includes not only synchrotron self-Compton
(SSC; e.g. \cite{Fan+08} and references therein) but also EIC emission
due to upscattering by external shock electrons \cite{Bel05,Pan07, Murase:2009su}.
Obtaining sufficient photon statistics for the time window of a single pulse in the prompt emission
would help to minimize such contributions from the afterglow,
which is expected to be less variable than the prompt emission
\cite{Murase:2009su,Ioka:2004gy}.

In addition to the break or cutoff due to $\tau_{\gamma\gamma}$,
the high-energy spectrum may reveal either a cooling break,
a maximum energy cutoff of synchrotron emission \cite{GG03,Ioka10,Wang:2009sd},
or a Klein-Nishina break of inverse Compton emission
\cite{GG03b,Bosnjak09,Nak+09}.
External $\gamma-\gamma$ absorption due to the EBL
is also unavoidable and can obscure an intrinsic break or cutoff (Section \ref{sec:EBL}). 
Detailed information on spectral variability from CTA
will be vital to discriminate between these possibilities.

\vspace{0.3cm}
\noindent{\bf Prompt Emission}.
Since the probability of serendipitously detecting a GRB in the FoV of CTA
during standard, pointing-mode observations is small,
rapid follow-up within their duration ($\sim 10$--$100\;$s for long GRBs)
is generally necessary to catch the prompt emission (see however, Section \ref{sec:widefield}).
Although the rate of GRB detection during the prompt phase is expected to be $\lesssim 1$ per year,
once it is achieved, hundreds or more photons
from a single event can be potentially obtained
and provide invaluable clues to the emission mechanism (Section \ref{sec:rates}).

The total energy radiated in the GeV band for LAT GRBs is typically a fair fraction
of that in the MeV band \cite{Beniamini11,Ackermann12}, and in some cases can be considerable.
The LAT spectra for some events remain hard up to the highest measured energies
\cite{090510-phys,090902B}, and the situation beyond remains unclear.
Robust estimates of the total radiated energy can impose constraints on the central engine.
For example, if the total intrinsic (i.e. collimation-corrected) energy is larger than a few $10^{52}\;$erg,
this might favor rapid accretion onto a newly born stellar-mass black hole \cite{Tho+04}
over a millisecond magnetar central engine for long GRBs,
although the collimation angle is usually poorly constrained on an event-by-event basis.
The energy budget is also an issue for hadronic models
in which the radiative efficiencies are generally limited (Section \ref{sec:UHECR}).

CTA is potentially capable of delineating the multi-GeV light curves
of GRBs with high photon statistics (Section \ref{sec:simulation-lc}),
which will be crucial to pin down the emission site of the high-energy photons. 
Very rapid and large amplitude variability would favor internal shocks
or photospheric emission over an external shock origin. Variability will also be the key
to discriminate leptonic or hadronic emission mechanisms,
since the acceleration and cooling timescales for protons and nuclei relevant
for the GeV-TeV emission are generically much longer than for leptons
(Section \ref{sec:UHECR}).

Although the delayed onset of the high-energy emission
may be generally challenging to address with CTA
as the typical delay $t_{\rm delay} \sim 1\;$s,
follow-up of a GRB possessing a precursor pulse might allow the observation of the onset of main burst phase.
Another possibility is prompt detection during wide-field mode observations (Section \ref{sec:widefield}).
Detailed characterization of the delay will be useful for
discriminating different models from the different dependences on key variables
such as luminosity $L$ and photon energy $\varepsilon_{\gamma}$.
For example,
if the delayed onset is caused by the evolution of the pair opacity cutoff, we expect
$t_{\rm delay} \propto L \ \varepsilon_{\gamma}$ for kinetic-energy dominated jets \cite{Li:2008ub},
and
$t_{\rm delay} \propto L^{0.14} \varepsilon_{\gamma}^{0.16}$
for magnetically-dominated jets \cite{BK12}.
Alternatively,
$t_{\rm delay} \propto L^{1/3} \varepsilon_{\gamma}$ if the delay is related to jet confinement \cite{Ioka11},
while
$t_{\rm delay}$ is independent of $L$ or  $\varepsilon_{\gamma}$
if it is determined by the size of the progenitor system in connection with baryon entrainment \cite{Ioka10,Toma11}.

Chances are higher for observing GRBs in coincidence with X-ray flares seen around $t \sim 100$--$10^5\;$s.
In view of their strong and rapid variability, they are usually attributed to
long-lasting activity of the central engine \cite{Ioka:2004gy,Burrows:2005ww},
although they may alternatively arise from sporadic late-time magnetic reconnection events
within a highly magnetized outflow \cite{Gian06}.
Investigating the high-energy properties of the X-ray flares in comparison with the prompt emission
can reveal their true origin.
{\it Fermi} detected a few GeV photons during the X-ray flaring activity of GRB 100728A \cite{Abdo2011XRF},
but cannot distinguish whether they originate from internal dissipation processes,
long-lasting afterglow emission, or moderately variable EIC emission
from external shock electrons \cite{Fan+08,Wan+06,GG08,Mur+11}.
High time resolution studies with the superior photon statistics of CTA will be crucial
to determine whether the high-energy photons and the X-ray flares are co-spatial,
as well as to constrain the bulk Lorentz factor of the emitting region and their emission mechanism.

The prompt emission of short GRBs may only be detectable by CTA through wide-field mode observations
(Section \ref{sec:widefield}).
However, it may still be possible to follow up the extended emission of short GRBs,
which is observed in the MeV band to last for $\sim 10^2\;$s and for which the total radiated energy
can sometimes be comparable to or even larger than that of the initial spike.
Although this emission is most likely related to the activity of the central engine \cite{Barthelmy:2005bx,Norris:2006rw},
its exact nature is unknown, and CTA observations could provide a valuable clue.

\vspace{0.3cm}
\noindent{\bf Afterglow Emission}.
A promising target for CTA follow-up is the high-energy afterglow emission of GRBs.
{\it Fermi} LAT has detected long-lived GeV emission up to $t \sim 10^3\;$s from dozens of events
and in some cases even up to $t \sim 4000-8000\;$s,
which is most likely related to the afterglow observed at lower energies \cite{090510-AG, Cenko11}.
There are good chances for CTA to detect such emission at higher energies (Section \ref{sec:rates}).

The known radio to X-ray afterglow emission is 
generally well explained as synchrotron radiation from electrons
accelerated in the external forward shocks resulting from the interaction
of the outflow with the ambient medium \cite{vanParadijs00},
especially at times later than $t \sim 10^3\;$s.
The simplest explanation of the GeV emission
as the high-energy extension of the synchrotron emission from an adiabatic blastwave
(together with suppression of the SSC emission)
requires rather extreme conditions such as a very low ambient density
and no magnetic field amplification \cite{KB-D09}.
The GeV light curve sometimes decays fairly rapidly as $\sim t^{-1.5}$,
which might be attributed to a radiative external shock,
although the ambient medium then needs to be enriched in
$e^{\pm}$ pairs at relatively large radii ($\sim 10^{16}\;$cm) \cite{Ghisellini10}.
Alternatively, the initial rapid decay may be related to inverse Compton cooling effects
in the Klein-Nishina regime \cite{Wan+10}, or 
perhaps more naturally, the gradual turn-off of the prompt emission
overlapping with the afterglow onset \cite{He+11,LW11,Max+11}.
An important test for the synchrotron afterglow model would be
detection of the maximum energy cutoff sweeping across
the observing band as the Lorentz factor declines \cite{PN10}.
In addition, at later times the Klein-Nishina suppressed SSC spectrum may become visible \cite{Wan+10}.
CTA could probe these features as well as the closure relation between the temporal and spectral indices
and thus help to determine the precise origin of the extended emission,
especially in combination with radio to X-ray multiwavelength observations.

CTA can also provide valuable information on the yet unclear nature
of the early X-ray afterglow at $t \lesssim 10^3\;$s,
particularly the shallow decay phase seen in most {\it Swift} bursts \cite{Nou+06,Obr+06,Zha+06},
which is not expected in standard models and whose origin has been debated \cite{Gra07}.
Possible models include late-time energy injection \cite{Zha+06,RM98,GK06},
long-lasting central engine activity \cite{Ghi+07,Kum+08},
time dependence of shock microphysics \cite{GKP06,Ioka:2005zj},
viewing angle effects \cite{Gra02,EG06}, shock coasting in a wind medium \cite{Shen12},
or contribution from SSC emission \cite{Petropoulou11}.
These predict different high-energy emission signatures (e.g. \cite{Fan+08,Mur+11})
so that simultaneous observations by CTA and X-ray instruments would be a key tool
to identify the actual mechanism.


\subsection{Ultra-High-Energy Cosmic Rays and Neutrinos}
\label{sec:UHECR}

If the prompt emission from GRBs results from electrons accelerated at internal shocks,
the energy in magnetic fields within GRB outflows is inferred to be a sizable fraction of the equipartition value.
Then protons and heavier nuclei are also likely to be accelerated under the same conditions,
potentially up to $\sim 10^{20}$~eV and beyond,
compatible with the observed energies of ultra-high-energy cosmic rays (UHECRs)~\cite{wax95,vie95,mil96}.
This may also apply to external shocks, particularly the reverse shock that occurs during the early afterglow \cite{Waxman00},
and possibly for the forward shock as well, albeit with nontrivial issues for acceleration to UHE \cite{Lemoine10}.
UHECR acceleration may also proceed via magnetic reconnection rather than shocks \cite{Giannios10}.
For GRBs to be viable sources of UHECRs,
their CR energy output must be consistent with the local UHECR emissivity,
$\varepsilon_{\rm CR}^2 d\dot{N}_{\rm CR}/d \varepsilon_{\rm CR}
\simeq (0.5-1) \times 10^{44}\ {\rm erg\ Mpc^{-3} yr^{-1}}$ at $10^{19}$~eV~\cite{wax98,mur09,kat09}.  
Compared with the estimated local GRB rate $\sim 0.1-1\ {\rm Gpc^{-3}~yr^{-1}}$~\cite{gue05,dai06,le07,Wanderman10},
the necessary isotropic-equivalent energy per burst in accelerated CRs 
$E_{\rm CR}^{\rm iso} \gtrsim 10^{54}$~erg
(depending somewhat on the assumed spectral index and the minimum CR energy $\gtrsim \Gamma A m_p c^2$,
where $A$ is the mass number of accelerated nuclei).
This generally exceeds the observed isotropic energy in MeV gamma-rays $E_\gamma^{\rm iso}$
by a large amount, entailing efficient CR acceleration as well as an energetic central engine.
It may point to significantly larger masses of black holes and/or accreted matter than commonly assumed
(note that black hole masses of $\gtrsim 30 M_\odot$ are inferred in some X-ray binaries \cite{Prestwich07}),
or that the outflow collimation angle is narrower than usually considered
so that the total intrinsic energy is actually not excessive.

If UHECRs are produced in internal shocks,
the above discussion also points to a high CR baryon-to-electron ratio
$E_{\rm CR}^{\rm iso}/E_{\rm e}^{\rm iso} \sim 1-10^{2.5}$~\cite{mur08,eic10,wax10},
as the energy in electrons $E_{\rm e}^{\rm iso}$ is considered to be comparable to $E_\gamma^{\rm iso}$.
Similar considerations are valid for an external shock origin of UHECRs,
implying that the isotropic kinetic energy of the blastwave is initially much larger than $E_\gamma^{\rm iso}$,
and either that only a small fraction of the electrons are accelerated \cite{Eichler05},
or that the CRs efficiently escape before the blastwave becomes adiabatic \cite{Dermer07}.
Such high proton-to-electron ratios in accelerated particles are observed in low-energy Galactic cosmic rays
and are also observationally inferred for the non-relativistic shocks of supernova remnants~\cite{aha06},
but the physical processes that determine the relative efficiency of acceleration and energy transfer
in collisionless astrophysical shocks are still poorly understood.
Dedicated multi-wavelength and multi-messenger observations are necessary to elucidate all these issues.

Efficient proton acceleration in GRBs may induce distinctive GeV-TeV band components in the prompt or afterglow emission
~\cite{vie97,boe98,tot98,Zhang01,Peer05,der06,gup07,asa07,asa09}.
Electron-positron pair cascades initiated by photomeson interactions between UHE protons and low-energy photons
can generate synchrotron and inverse Compton (IC) emission with a hard spectrum over a wide energy range from eV to TeV.
{\it Fermi} indeed detected hard spectral excess components above 0.1 GeV
from several bright GRBs in the prompt phase \cite{090510-phys,090902B,090926A}.
For GRB 090510~\cite{090510-phys} and 090902B~\cite{090902B},
low-energy spectral excesses over the Band component below tens of keV were also seen
that are consistent with extrapolations of the hard GeV spectra.
The hadronic pair cascade emission can readily explain
both the keV and GeV excesses \cite{Asano09,Asano10,mur12}.
In the case of GRB 090510,
the necessary isotropic-equivalent luminosity in accelerated protons
$L_{\rm p}^{\rm iso} \gtrsim 10^{55} \mbox{erg}~\mbox{s}^{-1}$~\cite{Asano09}
is much larger than the observed gamma-ray luminosity $L_{\gamma}^{\rm iso}$
and commensurate with the energetics requirements for the GRB-UHECR hypothesis.  
Models that attribute the hard GeV component to proton synchrotron radiation
also demand similarly large $L_{\rm p}^{\rm iso}$~\cite{Razzaque10}.
On the other hand,
the required value of $L_{\rm p}^{\rm iso}$ for GRB 090902B
is comparable to $L_{\gamma}^{\rm iso}$ ~\cite{Asano10},
so the fractions of energy carried by protons and electrons may vary from burst to burst. 

\begin{figure}[!htb]
\includegraphics[width=\columnwidth]{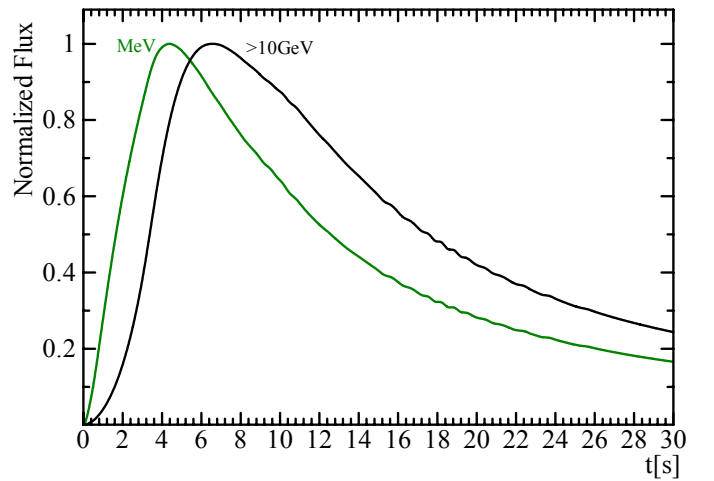}
\caption{
Comparison of light curves at MeV (green) and $>10$ GeV (black) of a single pulse
emitted from an expanding shell in the GRB outflow
for a hadronic photomeson-induced cascade model,
with bulk Lorentz factor $\Gamma=600$, initial emission radius $R_i=1.3 \times 10^{16}$ cm,
electron power $L_e=1.7 \times 10^{54}$ erg/s, proton/electron energy ratio $U_p/U_e=20$,
magnetic/electron energy ratio $U_B/U_e=3$, and redshift $z=4.35$ (see also \cite{Asano12}).
Realistic GRB light curves would comprise a superposed series of such pulses.
\label{fig:MeVGeVLC}}
\end{figure}

The limited statistics of multi-GeV photons for GRBs measured by {\it Fermi} LAT
does not allow us to distinguish hadronic models from alternative leptonic models
such as those based on synchrotron self-Compton \cite{bos09,cor10,asa11}
or external inverse Compton (EIC)~\cite{Bel05,tom09,Li:2008ub,Toma11} processes in the prompt phase,
or even the early afterglow emission~\cite{Ghisellini10,KB-D09}.
As most models fare reasonably well in reproducing the observed time-integrated (or coarsely time-resolved) spectra,
the key discriminant should be multiband variability data with high time resolution.
In the basic internal shock picture, each pulse in the MeV light curve is interpreted as synchrotron emission
from electrons in individual expanding shells corresponding to shocked regions within the GRB outflow,
and the width of the MeV pulse is mainly determined by the dynamical timescale of the shell \cite{Kobayashi97}.
The light curves at higher energies should reflect the nature of the emission mechanism.
For example, in one-zone SSC models, the GeV-TeV pulse width and shape should be similar to those at MeV
as the cooling timescales of the emitting electrons are typically much shorter than the dynamical time,
and the GeV light curve should closely track the MeV light curve
except for a marginal delay due to the finite buildup time of the seed photons.
Contrastingly, in hadronic models, the acceleration and cooling timescales of the highest-energy protons
that trigger the multi-GeV emission are comparable to the dynamical time.
While the low-energy and high-energy light curves should be broadly correlated on longer timescales
that reflect the dynamical history of the central engine,
the individual pulse profiles for the multi-GeV component are expected to be appreciably broader and smoother than at MeV.
This is clearly demonstrated from a detailed model calculation in Fig.\ref{fig:MeVGeVLC}
(see \cite{Asano12} for more details).
High photon statistics measurements of energy-dependent light curves by CTA (see Section \ref{sec:simulation-lc})
will provide a critical discriminant between leptonic and hadronic models,
as well as a unique and valuable test of the GRB origin of UHECRs.

The optical to X-ray bands may also be important for probing prompt UHECR generation in GRBs.
Besides the keV excess components seen in the bursts mentioned above, 
the bright, prompt optical emission of the ``naked-eye'' GRB 080319B \cite{rac08}
can also be interpreted as the lowest-energy portion of the hadronic cascade emission \cite{Asano10}.
Had this GRB occurred after the launch of {\it Fermi},
the accompanying high-energy emission could have been detectable.
Future searches for temporal correlations among the eV to multi-GeV bands
including rapid followup by robotic optical telescopes are desired.

For UHECR production in the afterglow,
the temporal behavior in specific energy bands
may not be sufficient to clearly decipher hadronic gamma-ray signatures.
However, unlike the prompt emission,
the radiation mechanism at lower frequencies is reasonably well understood as being due to accelerated electrons
(Section \ref{sec:GRBphys}).
This allows for comparatively robust predictions for the spectra and light curves of associated, leptonic high-energy components,
relative to which one can search for anomalous, hadronic components \cite{boe98,Zhang01,Peer05}.
Broadband coverage including CTA of the afterglow evolution over a range of timescales should provide a crucial diagnostic.
X-ray flares often seen during the early afterglow may also be potential sites of photohadronic gamma-ray emission
accompanying UHECR acceleration \cite{mur06a}, whose detection will be facilitated by its temporal signature.

Except for rare, nearby events,
gamma rays in the TeV regime may be difficult to detect
from classical long GRBs in view of the severe EBL attenuation
expected at these energies and their typical redshifts of $z \sim 1-2$.
However, low-luminosity GRBs occurring at much lower $z$
may constitute a separate class of transients and may possibly be more common,
with an estimated local rate of $\sim {10}^{2.5} \ {\rm Gpc^{-3}~yr^{-1}}$~\cite{sod06,lia07}.
Such events accompanied by hypernovae are also candidate sources of UHECRs~\cite{mur08,mur06b,wan07},
and the associated hadronic signals could be interesting targets for CTA~\cite{asa08,mur09b},
in addition to their high-energy afterglows~\cite{he09}.
For an event like GRB 060218 with $E_\gamma^{\rm iso} \sim {10}^{49}-{10}^{50}$~erg,
duration $T \sim 3000$~s, and distance $D \sim 140$~Mpc,
the estimated TeV flux is $\sim {10}^{-10.5}~{\rm TeV}~{\rm cm}^{-2}~{\rm s}^{-1}$
if $\sim 3$~\% of the proton energy is channeled to radiation~\cite{mur06b}.
This may be detectable by CTA even at TeV energies
if such low-$z$ events can be identified rapidly
and (at least part of) the MSTs can be slewed to achieve sufficient sensitivity.

Recent observations of UHECRs by the Pierre Auger Observatory in the southern hemisphere
give hints that their composition may be dominated by heavy nuclei rather than protons
at the highest energies~\cite{abr10} (see however \cite{abb10,tsu11}
for contrasting results from HiRes and Telescope Array in the northern hemisphere).
GRBs should also be able to accelerate heavy nuclei to ultra-high energies,
but their survival against internal photodisintegration processes
mandates photomeson interactions to be inefficient~\cite{mur08,wan08,met11},
so that $\lesssim 0.1$~\% of the cosmic-ray energy can be converted into cascade gamma-rays and neutrinos.
Nevertheless, as internal $\gamma\gamma$ attenuation is then mitigated~\cite{mur08},
GeV-TeV signals from nuclear de-excitation, photopair creation or nuclear synchrotron emission
may be observable and provide a unique signature of UHECR nuclei acceleration~\cite{mur10b,Inoue12}.
For example, Lorentz-boosted de-excitation gamma rays
at energies $\sim {\rm TeV} (E_A/3 \times {10}^{16}~{\rm eV})$, where $E_A$ is the energy of CR nuclei,
may be detectable from nearby, low-luminosity GRBs (see however \cite{AhaTay10}).

The photomeson interactions described above will also give rise to high-energy neutrinos,
primarily at multi-TeV to PeV energies, whose detection will provide definitive evidence of
hadron acceleration in GRBs~\cite{wax97,der03,mur06}.
Ongoing measurements by the IceCube neutrino observatory, mainly sensitive to sources in the northern sky,
have begun to provide meaningful constraints on theoretical predictions.
Stacking analyses of data in coincidence with GRBs indicate that
the average photomeson production efficiency during the prompt phase is not very high \cite{abb11,abb12}.
However, the current predictions involve large uncertainties,
and it is too early to entirely reject the GRB-UHECR scenario at this moment.
As it is difficult to obtain detailed information for individual bursts through neutrinos alone,
gamma-ray observations will play an independent and complementary role
in probing the GRB origin of UHECRs.
In the future, synergy can also be expected with the KM3NeT facility
that will be more sensitive for neutrino sources in the southern sky \cite{Sapienza11},
as well as the Askaryan Radio Array that is better suited for the EeV energy range \cite{Allison12}.

The UHECRs themselves are expected to arrive at us with significant time delays of up to $\Delta t \sim 10^7$ yr
due to deflections in intergalactic magnetic fields (IGMFs) during their propagation \cite{Miralda96},
long after the emission from the burst itself has faded away.
Thus UHECR observations alone are unable to directly identify the sources in the case of GRBs,
making gamma-ray and neutrino observations indispensable for this purpose.
After the UHECRs have escaped from their sources, they can experience
further photomeson interactions with the CMB and/or EBL during their propagation
and induce secondary cascade gamma rays that are delayed and/or spatially extended,
either through inverse Compton \cite{WaxCop96} or synchrotron \cite{Aharonian10} processes.
Such emission may be detectable and would provide a further unique probe of UHECRs,
although its properties depend sensitively on the highly uncertain strength and structure of IGMFs.


\subsection{Extragalactic Background Light and Intergalactic Magnetic Fields}
\label{sec:EBL}

The extragalactic background light (EBL) refers to the diffuse and nearly isotropic background
of infrared-optical-ultraviolet radiation originating outside of our Galaxy.
Although its principal source is thought to be emission from the ensemble of
normal stars in the Universe integrated throughout its history, 
other objects such as quasars or Population III stars may also provide important contributions,
as well as possibly exotic processes such as dark matter decay.
Since the EBL embodies fundamental information for studies of cosmology and galaxy formation,
understanding its detailed properties is of paramount importance.
However, its amplitude, spectrum and evolution are difficult to determine precisely from direct measurements, 
in particular because of the strong but uncertain contamination from zodical foreground emission.

An indirect but powerful means of probing the EBL
is through photon-photon ($\gamma\gamma$) absorption of high-energy gamma-rays.
Gamma-rays emitted from extragalactic sources
can be absorbed during intergalactic propagation by interacting
with photons of the EBL
to produce electron-positron pairs,
as long as there is sufficient opacity to the process.
The observed spectra of the gamma-ray sources should then exhibit corresponding attenuation features,
from which one can deduce or constrain the properties of the EBL in a redshift-dependent way.
This method has been effectively utilized in TeV observations of blazars
by current ground-based Cherenkov telescopes
to set important constraints on the EBL in the near infrared to optical bands
at relatively low redshifts, $z \lesssim 0.5$ \cite{Aharonian06,Albert08}
(see also \cite{Krennrich12,Mazin12} for reviews).

GRBs are the most luminous and distant gamma-ray emitting objects known in the Universe,
typically arising at redshifts $z \sim 1-4$,
which correspond to the peak epoch of cosmic star formation activity \cite{Butler10}.
Furthermore, they are known to occur at least up to $z \sim 8-9$ \cite{Tanvir09,Salvaterra09,Cucchiara11},
well into the cosmic reionization era,
and possibly even beyond, out to the very first epochs of star formation in the Universe \cite{Bromm06}
(see \cite{TomaP3} and references therein for the possibility of especially luminous emission by GRBs
from Pop III progenitors).
The recent detections by {\it Fermi} LAT of dozens of GRBs
including GRB 080916C at $z = 4.35$ \cite{080916C}
clearly demonstrate that at least some GRBs have luminous emission extending to few tens of GeV
(corresponding to rest-frame energies of up to $\sim$ 100 GeV),
and the duration of the multi-GeV emission can last up to several thousand seconds.
Thus there are good prospects for CTA of providing a powerful probe of the EBL
through the multi-GeV spectra of GRBs with high quality at $z > 1$,
as well as for GRBs at $z \gtrsim 3$ and beyond, a regime that cannot be explored with active galactic nuclei (AGNs) \cite{Sol12}.

Direct observational determination of the cosmic star formation rate at high redshifts
is plagued by various uncertainties,
particularly the contribution of faint galaxies below the detection limit of optical-IR telescopes.
An in-situ probe of the EBL utilizing gamma-ray absorption in high-$z$ GRBs can circumvent this problem.
At $z>1$, the EBL waveband relevant for gamma-ray absorption moves into the UV,
providing a more direct measure of the contribution from massive stars.
Current theoretical predictions at these redshifts can differ quite strongly among different models, e.g.
\cite{Kneiske04,Stecker06,Franceschini08,Gilmore09,Finke10,Dominguez11,Gilmore12,YInoue12}.
Although {\it Fermi} detections of AGNs and GRBs up to $z=4.35$
have provided important upper limits to the EBL in this redshift range \cite{Abdo10},
the associated spectral cutoffs have not been measurable because of the limited photon statistics.
The much larger statistics expected for GRB detections by CTA
(a few tens to hundreds of photons above 30 GeV for typical events and even more for brighter events, see Section \ref{sec:rates})
should allow more robust measurements of EBL-induced cutoffs,
leading to more reliable determinations of the EBL and thus the total cosmic star formation at high redshifts.

In the redshift range $z \sim 3-4$,
various observations indicate that HeII (singly ionized helium) in the intergalactic medium
was reionized \cite{Jakobsen94}, although the details are not yet understood.
Since this process requires photons with energy $>$54.4 eV and is difficult to attain with stars alone, 
the implication is that objects with hard UV spectra, most likely quasars,
make an important contribution to the UV EBL at these redshifts
(e.g. \cite{Furlanetto08,Faucher08,Faucher09} and references therein).
EBL absorption measurements in $z \gtrsim 3$ GRBs
may thus offer invaluable insight into the reionization of intergalactic HeII
as well as the global history of accretion onto supermassive black holes at these epochs,
in addition to cosmic star formation.

An exciting, albeit challenging possibility is the detection of GRBs at $z \gtrsim 6$ in the cosmic reionization era.
Some time after the epoch of cosmic recombination at redshift $z \sim 1100$,
the bulk of the intergalactic hydrogen in the universe must have been somehow reionized by $z \sim 6$,
as indicated observationally from the spectra of high-$z$ quasars and the polarization of the CMB.
However, the sources, history and nature of this cosmic reionization process are still largely unknown,
as most of this redshift range has yet to be explored through direct observations.
Because the first stars and galaxies in the universe must have formed during this period,
the primary suspect is photoionization by UV radiation from such objects,
potentially involving metal-free, Population III stars.
Alternative possibilities include mini-quasars, supernova remnants and dark matter decay.
Besides providing us with clues to such processes in the early universe,
cosmic reionization also profoundly affects the ensuing formation of stars and galaxies,
so elucidating this era is one of the most pressing issues in observational cosmology today
(for reviews, see e.g. \cite{Barkana01,Ciardi05}).
As discussed by a number of authors \cite{Oh01,SInoue10,YInoue12},
UV radiation fields with sufficient intensities to cause the reionization of the intergalactic medium (IGM)
are also likely to induce appreciable gamma-ray absorption in sources at $z \gtrsim 6$
at observed energies in the multi-GeV range,
with a potentially important contribution from Pop III stars.
Measurements of these effects can thus provide
important cross-checks of current models of cosmic reionization,
a unique probe of the evolving UV EBL
during the era of early star formation,
as well as a test for the existence of the yet hypothetical Pop III stars.
The detection of such very high-$z$ GRBs by CTA may be quite rare, however (Section \ref{sec:rates}).

A generic issue for EBL absorption studies
is distinguishing them from spectral cutoffs intrinsic to the source.
In this regard, the spectral variability inherent in GRBs offers an advantage.
In general, changes in physical conditions of the source
that cause variations in flux should also be accompanied
by variations of the intrinsic cutoff energy,
whether they are due to injection of freshly accelerated particles,
changes in the magnetic fields, internal radiation fields, bulk flow velocity, etc.
In contrast, cutoffs of EBL origin should be stable in time
and independent of the variability state of each object.
Acquisition of time-resolved spectra by CTA
should thus greatly help in the deconvolution of the two effects.

The $\gamma\gamma$ interactions between primary gamma rays from the GRB and low-energy photons of the EBL
generate electron-positron pairs far away from the source, typically inside intergalactic void regions.
The pairs can then be deflected by weak intergalactic magnetic fields (IGMFs) for a short period within the voids
before giving rise to secondary GeV-TeV components by upscattering ambient CMB photons.
Known as ``pair echos'', they arrive with a characteristic time delay relative to the primary emission
that depend on the properties of the IGMF and hence constitute a valuable probe of their nature
\cite{Plaga95,Razzaque04,Ichiki08,Takahashi08,Murase09,Takahashi11}.
For GRBs at $z \lesssim 1$,
the pair echos can last longer than the prompt emission and be detectable during the afterglow phase
as long as the amplitude of IGMFs in voids are $B_{\rm IG}\lesssim 10^{-18}$ G
and the primary GRB spectra extend into the multi-TeV regime~\cite{Ichiki08,Takahashi08,Murase09}.
They can be distinguished from the high-energy afterglow
through their typically hard spectra and exponentially decaying light curves.
Pair echos from GRBs at higher redshifts are potentially sensitive to somewhat stronger IGMFs,
although their detectability demands more extreme properties for the primary emission~\cite{Takahashi11}
\footnote{For blazar AGNs, the time-integrated energy of the primary emission can exceed that of GRBs
so that for a certain range of IGMF strengths, the secondary emission may be observable as spatially extended halos
rather than through their time delay \cite{Neronov09,Sol12}.}.
The detection of pair echos or even upper limits to such components
will provide unique and valuable constraints on IGMFs, whose nature is currently poorly understood
but may be related to physical processes in the early Universe or the cosmic reionization epoch \cite{Widrow02}.


\subsection{Lorentz Invariance Violation}
\label{sec:LIV}

Some models of quantum gravity (QG) allow violation of Lorentz
invariance, and in particular allow the photon propagation speed
$v_{\rm ph}$ to depend on its energy $E_{\rm ph}$: $v_{\rm
ph}(E_{\rm ph}) \neq c$, where $c\equiv\displaystyle\lim_{E_{\rm
ph}\to 0}v_{\rm ph}(E_{\rm ph})$. The Lorentz invariance violating
(LIV) part in the dependence of the photon momentum $p_{\rm ph}$
on its energy $E_{\rm ph}$ 
can be expressed as a power series,
\begin{equation}\label{eq:QG1}
\frac{p_{\rm ph}^2c^2}{E_{\rm ph}^2}-1 
= \sum_{k = 1}^\infty s_k\left(\frac{E_{\rm ph}}{M_{{\rm QG},k}c^2}\right)^k\ ,
\end{equation}
in the ratio of $E_{\rm ph}$ and a typical energy scale $M_{{\rm
QG},k}c^2$ for the $k^{\rm th}$ order, which is expected to be up to
the order of the Planck scale $M_{\rm Planck} = (\hbar c/G)^{1/2}
\approx 1.22\times 10^{19}\;{\rm GeV/c^2}$, where $s_k \in
\{-1,\,0,\,1\}$. Since we observe photons of energy well below the
Planck scale, the dominant LIV term is associated with the lowest
order non-zero term in the sum, of order $n = \min\{k|s_k\neq 0\}$,
which is usually assumed to be either linear ($n = 1$) or quadratic
($n = 2$). The photon propagation speed is given by the corresponding
group velocity
\begin{equation}
v_{\rm ph} = \frac{\partial E_{\rm ph}}{\partial p_{\rm ph}} \approx
c\left[1-s_n\,\frac{n+1}{2}\left(\frac{E_{\rm ph}}{M_{{\rm QG},n}c^2}\right)^n\,\right]\ ,
\end{equation}
where $s_n = 1$ corresponds to the sub-luminal case ($v_{\rm ph} <
c$ and a positive time delay), while $s_n = -1$ corresponds to the
super-luminal case ($v_{\rm ph} > c$ and a negative time delay).
Taking into account cosmological effects~\cite{JP08}, this induces a
time delay or lag in the arrival of a high-energy photon of energy
$E_{\rm h}$, compared to a low-energy photon of energy $E_{\rm l}$
emitted simultaneously from the same location, of
\begin{equation}\label{eq:Dt}
\Delta t = s_n\,\frac{(1+n)}{2H_0}\frac{\left(E_h^n-E_l^n\right)}{(M_{{\rm
QG},n}c^2)^n} \int_{0}^{z}\frac{(1+z^{\prime})^n}
{\sqrt{\Omega_m(1+z^{\prime})^3+\Omega_{\Lambda}}}\,dz^{\prime}\ .
\end{equation}
The focus here is on {\it Fermi} results for a linear energy dependence ($n
= 1$), which are the best to date from direct time of flight
measurements.

Applying Eq.~(\ref{eq:Dt}) to the highest energy photon detected in
GRB~080916C, of energy $E_{\rm h} = 13.22^{+0.70}_{-1.54}\;$GeV, which
arrived at $t = 16.54\;$s after the GRB trigger (i.e. the onset of the
$E_l \sim 0.1\;$MeV emission) resulted in a limit of $M_{{\rm QG},1} >
0.11M_{\rm Planck}$ for the sub-luminal case ($s_n = 1$) , when making
the conservative assumption that it was emitted anytime after the GRB
trigger (or the onset of any observed emission from this GRB)
\cite{080916C}, i.e. $\Delta t \leq t$. This avoids the need to
associate the highest energy photon with a particular spike in the
low-energy light curve, which is hard to do in a very robust way.
This limit was the strictest of its kind at that time.

\begin{figure}[!htb]
\centerline{\hspace{0.88cm}\includegraphics[width=0.9\columnwidth]{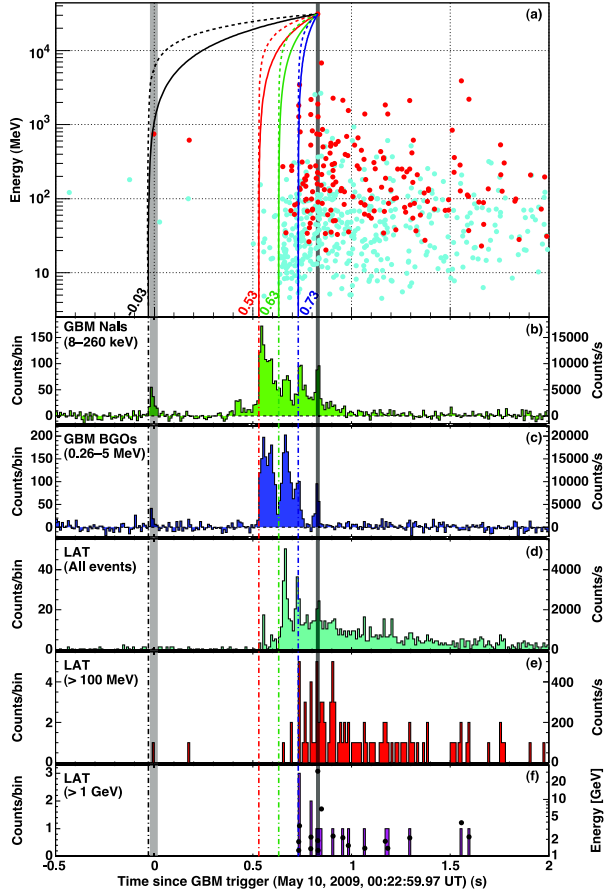}}
\caption{
Light curves of GRB~090510 at different energies. For details see text and \cite{090510-LIV}.}
\label{fig:090510-LIV}
\end{figure}

However, the next very bright LAT GRB, 090510, was short and had very
narrow sharp spikes in its light curve (see Fig~\ref{fig:090510-LIV}),
thus enabling to set even better limits~\cite{090510-LIV}. The main
results for GRB~090510 are summarized in Table~\ref{tab:LIV}. The
first 4 limits are based on a similar method as described above for
GRB~080916C, using the highest energy photon, $E_h =
30.53_{-2.56}^{+5.79}\;$GeV, and assuming that its emission time $t_h$
was after the start of a relevant lower energy emission episode: $t_h
> t_{\rm start}$. These 4 limits correspond to different choices of
$t_{\rm start}$, which are shown by the vertical lines in
Fig.~\ref{fig:090510-LIV}. The low end of the $1\;\sigma$ confidence
interval for the highest energy photon ($E_h = 28\;$GeV) and for the
redshift ($z = 0.900$) were used for conservativeness. The most
conservative assumption of this type is associating $t_{\rm start}$
with the onset of any detectable emission from GRB~090510, namely the
start of the small precursor that GBM triggered on, leading to $\xi_1
= M_{{\rm QG},1}/M_{\rm Planck} > 1.19$. However, it is highly
unlikely that the $31\;$GeV photon is indeed associated with the small
precursor. It is much more likely associated with the main soft
gamma-ray emission, leading to $\xi_1 > 3.42$. Moreover, for any
reasonable emission spectrum, the $31\;$GeV photon
would be accompanied by a large number of lower energy
photons, which would suffer a much smaller time delay due to LIV
effects, and would therefore mark its emission time.
Such photons with energies above $\sim 100\;$MeV could easily be detected by {\it Fermi} LAT, and therefore the
fact that significant high-energy emission is observed only at later
times (see Fig.~\ref{fig:090510-LIV}) strongly indicates that the
$31\;$GeV photon was not emitted before the onset of the observed
high-energy emission. One could choose either the onset time of the
emission above $100\;$MeV or above $1\;$GeV, which correspond to
$\xi_1 > 5.12$, and $\xi_1 > 10.0$, respectively.~\footnote{Note
that there is no evidence for LIV induced energy dispersion that might
be expected if indeed the $31\;$GeV photon was emitted near our
choices for $t_{\rm start}$ together with lower energy photons,
for any reasonable emission spectrum. This is evident
from the lack of accumulation of photons along the {\it solid} curves
in panel (a) of Fig.~\ref{fig:090510-LIV}, at least for the first 3
$t_{\rm start}$ values, and provides support for these choices of
$t_{\rm start}$ (i.e. that they can indeed serve as upper limits on a
LIV induced energy dispersion).}

The 5$^{\rm th}$ and 6$^{\rm th}$ limits in Table~\ref{tab:LIV} are
more speculative, as they rely on the association of an individual
high-energy photon with a particular spike in the low-energy light
curve, on top of which it arrives. While these associations are not
very secure (the chance probability is roughly $\sim 5-10\%$), they
are still most likely, making the corresponding limits interesting,
while keeping this big caveat in mind. The allowed emission time of
these two high-energy photons, if these associations are real, is
shown by the two thin vertical shaded regions in
Fig.~\ref{fig:090510-LIV}. For the $31\;$GeV photon this gives a limit
of $\xi_1 > 102$ for either sign of $s_n$.

The last limit in Table~\ref{tab:LIV} is based on a different method,
which is complementary and constrains both signs of $s_n$. It relies
on the highly variable high-energy light curve, with sharp narrow
spikes, which would be smeared out if there was too much energy
dispersion, of either sign. We have used the DisCan
method~\cite{DisCan} to search for linear energy dispersion in the
LAT data within the photon energy range 35 MeV - 31 GeV
~\footnote{We obtain similar results even if we use only
photons below $3\;$GeV or $1\;$GeV.} during the most intense emission
interval (0.5$\,$--$\,$1.45$\,$s). This approach extracts dispersion
information from all detected LAT photons and does not involve binning
in time or energy. Using this method we obtained a robust lower limit
of $\xi_1 > 1.22$ (at the 99\% confidence level). 

The most conservative limits (the first and last limits in
Table~~\ref{tab:LIV}) rely on very different and largely independent
analysis, yet still give a very similar limit: $M_{{\rm QG},1} >
1.2M_{\rm Planck}$. This lends considerable support to this result,
and makes it more robust and secure than for each of the methods
separately.

{\it Swift} found a separate, small precursor $\sim 13\;$s before the GBM precursor
that marked the trigger time of GRB~090510 \cite{TRG10},
itself about half a second before the start of the main GRB activity.
It has been claimed that this can significantly affect the {\it Fermi} limit derived under the assumption that the
31~GeV photon was emitted after the onset of any observed emission from GRB~090510.
However, it is highly unlikely that the
31~GeV photon was indeed emitted during this {\it Swift} precursor since (1)
as discussed above for the GBM precursor,
it should have been accompanied by a large number of lower energy photons
that were in fact not observed,
and (2) fine tuning is required for the 31~GeV photon to arrive on top of the brightest
emission episode (and also exactly on top of a bright and narrow spike seen at all energies).
Therefore, the {\it Swift} precursor is unlikely to have major consequences for the derived LIV limit.

{\bf LIV Prospects for CTA.}
In the conventional mode of IACT observations via followup of GRB alerts,
improving on the limit set by {\it Fermi} for GRB~090510
from the arrival time of the highest energy photon
would require detection of a GRB at a similar redshift ($z \sim 1$)
at energies $E_h \gtrsim 1(T_{\rm delay}/30\;{\rm s})\;$TeV,
where $T_{\rm delay}$ is the response time from the GRB trigger to the start of CTA observations.
This would be quite challenging,
since $T_{\rm delay}\gtrsim 30\;$s is generally expected (Sections \ref{sec:status-VHE}, \ref{sec:alert}),
and attenuation by the EBL is also likely to limit the detectability of photons from $z \gtrsim 1$ to $E_h \lesssim 1\;$TeV
\cite{Krennrich12,Mazin12}.

In contrast, for the DisCan method that searches for energy dispersion effects in the light curve,
the much larger effective area of CTA compared to {\it Fermi} LAT
(by roughly 4 orders of magnitude at $30\;$GeV \cite{Funk12})
can greatly enhance the photon statistics and allow significant progress,
as long as sufficiently variable emission occurs at the energies and timescales relevant for CTA.
For example, observing a large number of $\sim 0.1\;$TeV photons
with strong variability on timescales $t_v\sim 0.1\;$s
from a GRB at $z\sim 1$ (Section \ref{sec:simulation-lc})
might improve the {\it Fermi} limit by a factor of $\sim 30$.
At least at lower energies, such variability is seen in some long duration GRBs
tens of seconds after the GRB onset, within the CTA response time.

If a short GRB can be observed from its very onset during the wide-field mode of CTA observations (Section \ref{sec:widefield}),
a considerable number of $\sim 0.1-1\;$TeV photons with variability timescale of a few seconds
may be detectable,
potentially improving the {\it Fermi} limit by up to a factor of $\sim 10^3$.
Thus, such wide-field mode observations
can make a profound impact on fundamental physics by probing LIV with extraordinary precision.

All methods of constraining LIV require large amplitude, short timescale variability and bright high-energy emission.
Hence the prompt emission is more favorable for this purpose than the afterglow,
which is fainter and generally has a smooth temporal profile.
Nevertheless, the X-ray flares that are often observed by Swift superimposed on the afterglow
may also have correlated emission at GeV-TeV energies, as predicted in some models \cite{Wan+06,GG08,Fan+08}
and may have been seen in GRB 100728A \cite{Abdo2011XRF}.
In view of the higher expected detection rates in the afterglow phase with CTA (Section \ref{sec:rates}),
such late-time flares may also be interesting for probing LIV.
However, most X-ray flares have durations $\Delta t \gtrsim 0.1t$ with respect to the post-trigger time $t$
(possibly being a factor of $\sim 2$ longer at GeV-TeV \cite{Fan+08}),
and their luminosities rapidly decrease with $t$ \cite{Margutti11}.
Compared to the prompt emission for which the average duration and luminosity of spikes in the light curve are roughly constant,
LIV constraints from high-energy flares during the afterglow are thus expected to be weaker.

\newcommand{\rb}[1]{\raisebox{1.3ex}[0pt]{#1}}
\newcommand{\lb}[1]{\raisebox{-1.3ex}[0pt]{#1}}
\begin{table}[h!]
\begin{center}
{\footnotesize
\begin{tabular}{|c|c|c|c|c|c|}
\hline
$t_{\rm start}$ & limit on          & Reason for choice of                   & 
$E_l$ &     & limit  \\ 
(ms)            & $|\Delta t|$ (ms) & $t_{\rm start}$ or limit on $\Delta t$ & 
(MeV) & \rb{$s_n$} & on $\xi_1$
\\
\hline
\hline
$-30$ & $<859$ & {\scriptsize start of any observed emission}  & 
0.1  & 1 & $>$1.19 \\
$530$ & $<299$ & {\scriptsize start of main $<\,$1$\,$MeV emission} & 
0.1  & 1 & $>$3.42 \\
$630$ & $<199$ & {\scriptsize start of $>100\;$MeV emission}    & 
100  & 1 & $>$5.12 \\
$730$ &  $<99$ & {\scriptsize start of $>1\;$GeV emission}      & 
1000 & 1 & $>$10.0 \\
\hline
 ---  &  $<10$ & {\scriptsize association with $<\,$1$\,$MeV spike} & 
0.1  & $\pm\,$1 & $>$102  \\
 ---  &  $<19$ & {\scriptsize if $0.75\,$GeV $\gamma$ is from $1^{\rm st}$ spike} & 
0.1  & $\pm\,$1 & $>$1.33  \\ \hline
\multicolumn{2}{|c|}{$|\frac{\Delta t^{^{}}}{\Delta E_{}}| < 30\;\frac{\rm ms}{\rm GeV}$} & 
{\scriptsize lag analysis of all LAT events} &
--- & $\pm\,$1 & $>$1.22
\\
\hline 
\end{tabular}}
\end{center}
\caption{\footnotesize
Lower-limits on the Quantum Gravity (QG) mass scale associated with a
possible linear ($n=1$) variation of the speed of light with photon
energy, $M_{\rm QG,1} = \xi_1 M_{\rm Planck}$, that can be placed from
the lack of time delay (of sign $s_n$) in the arrival of high-energy
photons relative to low-energy photons, from {\it Fermi} LAT and GBM
observations of GRB~090510. See \cite{090510-LIV} for more details.
\label{tab:LIV}}
\end{table}


\section{Simulations of GRB Observations}
\label{sec:simulation}

In order to quantify the prospects for CTA observations,
we now present some simulated spectra and light curves of GRBs.
Although our ultimate aim is to assess the different science cases discussed above,
in view of the wide range of uncertainties in the current physical models,
here we take a purely phenomenological approach as a first step.
Choosing as templates some prominent bursts
detected by {\it Fermi} LAT whose spectra and variability
were relatively well characterized up to multi-GeV energies,
we simply assume that their intrinsic spectra extend to higher energies
as power-law extrapolations,
while accounting for the effects of EBL attenuation based on selected models.
These simulations should be considered exemplary first results
on which we can elaborate further in the future
by incorporating more physical ingredients
depending on the specific science motivation.

\subsection{Simulations of spectra}
\label{sec:simulation-spec}

A series of spectral simulations have been conducted
utilizing version 4 of the CTA simulation tool developed by Daniel Mazin and colleagues
(see \cite{Bernloehr12} for more details).
Our assumptions are as follows.

\begin{itemize}

\item Take as templates the bright LAT GRBs 090902B ($z=1.822$) \cite{090902B} and 080916C ($z=4.35$) \cite{080916C}.

\item Extrapolate their spectra to higher energies using the spectral indices measured by LAT at specific time intervals.

\item Using the time decay indices measured by LAT, normalize the flux at some post-trigger time $t_0$,
allowing for some delay in the telescope response
(sometimes optimistically, e.g. 35 sec, otherwise quite reasonably, e.g. 50-150 sec).

\item Besides the template bursts with their actual redshifts,
consider also events with the same intrinsic properties
but with redshifts scaled to different values from $z=1$ to 6.5,
accounting for spectral and temporal redshift corrections.

\item Adopt a range of EBL models, e.g. \cite{Kneiske04, Franceschini08, Finke10, SInoue10, YInoue12}.

\item Generally take array configuration E (perceived as a balanced choice for a broad range of science goals),
but also configuration B (with optimal performance for the lowest energies) in a few selected cases.
(See \cite{Bernloehr12} for more information on array configurations.)

\item Assume 20 deg for the zenith angle of observation.

\item Simulate the spectra that would be measured by CTA with the aforementioned tool,
taking exposure times in accord with the considered $t_0$.

\item For GRB 090902B, the considered $t_0$ are all in the extended emission phase at $t_0>$25 sec,
and we take photon index $\Gamma=-2.1$ and time decay index $\delta_t=-1.5$ \cite{090902B}.

\item For GRB 080916C, the considered $t_0$ correspond to time interval ``d'' ($t_0=$16-55 sec for $z=4.3$)
with $\Gamma=-1.85$.
Note that this spectral index is from the LAT only fits and not the GBM+LAT joint fits.
The time decay index is always $\delta_t=-1.2$.

\item As the template bursts are at the upper end of the luminosity distribution
and the probabililty of their detection could be relatively low (Section \ref{sec:rates}), 
consider also events with the same intrinsic properties
but with fluxes scaled by factor 1/10
that may correspond to bursts with more typical luminosities.

\end{itemize}

\begin{figure}[!htb]
\centerline{\hspace{0cm}\vspace{0cm}\includegraphics[width=1\columnwidth]{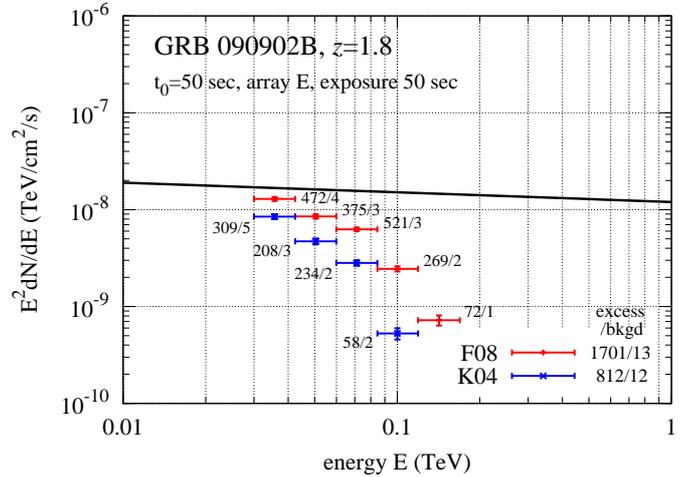}}
\caption{
Simulated spectra for GRB 090902B at $z=$1.8, for exposure time 50 sec and array configuration E,
adopting the EBL model of Franceschini et al. \cite{Franceschini08} (red)
and the ``best fit'' EBL model of Kneiske et al. \cite{Kneiske04} (blue).
The number of detected photons/background events are denoted for each energy bin beside the data points,
as well as for all energies in the legend.
The assumed source flux is $dN/dE=1.2 \times 10^{-8} (E/{\rm TeV})^{-2.1} {\rm cm^{-2} s^{-1} TeV^{-1}}$ (black solid line),
representing a power-law extrapolation of the observed {\it Fermi} LAT spectrum at $t=$50 sec after trigger.
}
\label{fig:GRB090902B_z1.8}
\end{figure}

\begin{figure}[!htb]
\centerline{\hspace{0cm}\vspace{0cm}\includegraphics[width=1\columnwidth]{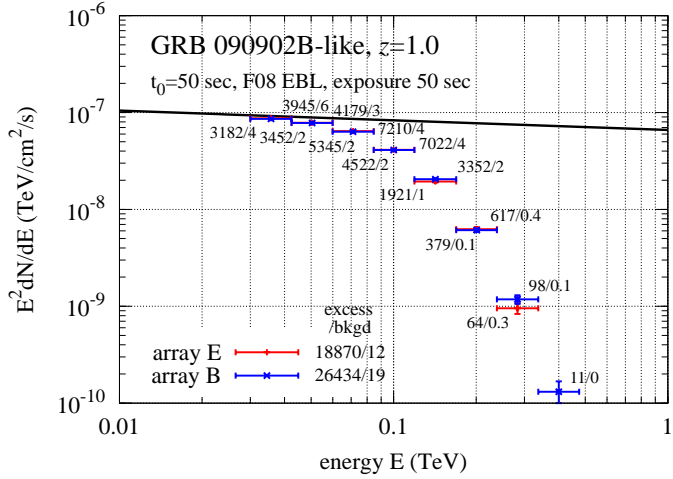}}
\caption{
Simulated spectra for a GRB 090902B-like event scaled to $z=$1.0, for exposure time 50 sec
and array configurations E (red) and B (blue), adopting the EBL model of Franceschini et al. \cite{Franceschini08}.
The number of detected photons/background events are denoted for each energy bin beside the data points,
as well as for all energies in the legend.
The assumed source flux is $dN/dE=6.6 \times 10^{-8} (E/{\rm TeV})^{-2.1} {\rm cm^{-2} s^{-1} TeV^{-1}}$ (black solid line),
representing a power-law extrapolation of the {\it Fermi} LAT spectrum of GRB 090902B
scaled to $z=$1.0 and $t=$50 sec after trigger.
}
\label{fig:GRB090902B_z1.0}
\end{figure}

\begin{figure}[!htb]
\centerline{\hspace{0cm}\vspace{0cm}\includegraphics[width=1\columnwidth]{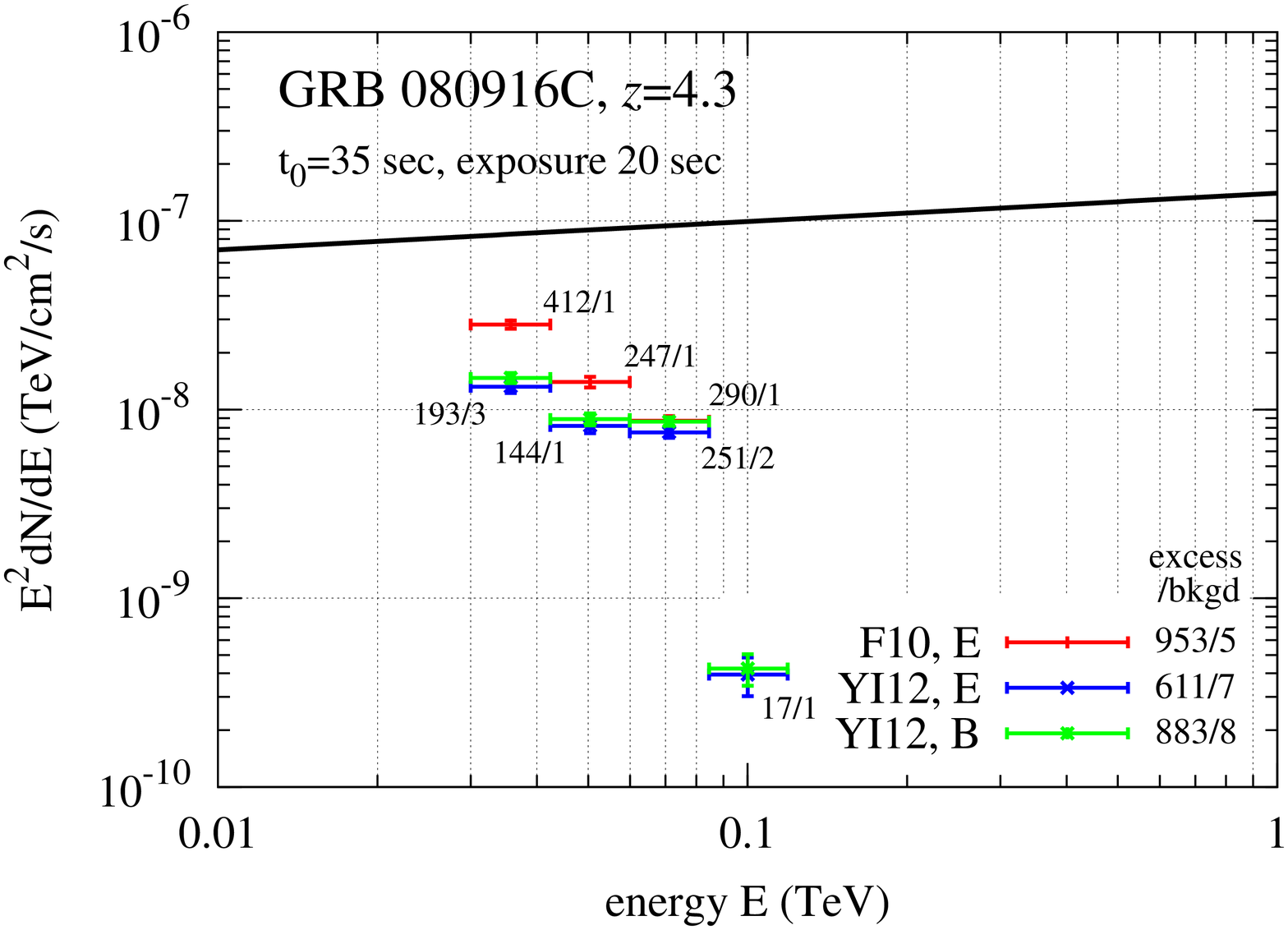}}
\caption{
Simulated spectra for GRB 080916C at $z=$4.3 for exposure time 20 sec and array configuration E,
adopting the EBL models of Finke et al. \cite{Finke10} (red) and Y. Inoue et al. \cite{YInoue12} (blue).
Also shown is the case of array configuration B for the latter EBL model (green).
The number of detected photons/background events are denoted for each energy bin beside the data points,
as well as for all energies in the legend.
The assumed source flux is $dN/dE=1.4 \times 10^{-7} (E/{\rm TeV})^{-1.85} {\rm cm^{-2} s^{-1} TeV^{-1}}$ (black solid line),
representing a power-law extrapolation of the observed {\it Fermi} LAT spectrum at $t=$35 sec after trigger (interval d).
}
\label{fig:GRB080916C_z4.3}
\end{figure}

\begin{figure}[!htb]
\centerline{\hspace{0cm}\vspace{0cm}\includegraphics[width=1\columnwidth]{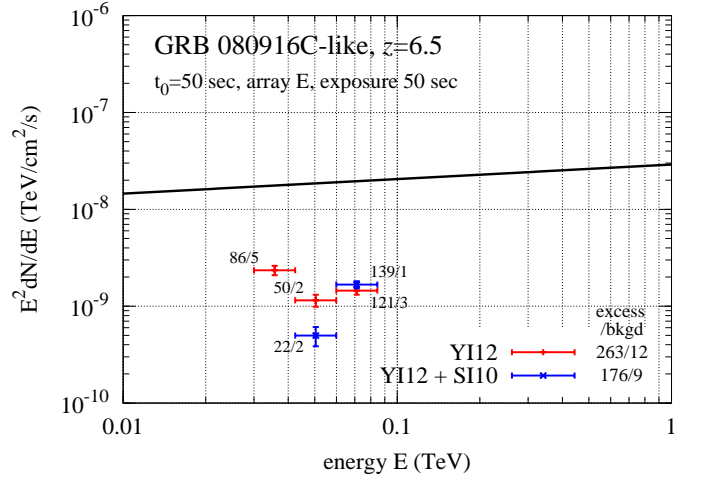}}
\caption{
Simulated spectra for a GRB 080916C-like event scaled to $z=$6.5, for exposure time 50 sec and array configuration E,
adopting the EBL model of Y. Inoue et al. \cite{YInoue12} (red), 
compared with that of S. Inoue et al. \cite{SInoue10} for $z>5$ in combination with Y. Inoue et al. for $z<5$ (blue;
note that the spectral shape rising sharply with energy
is an unrealistic artifact of combining these different models).
The number of detected photons/background events are denoted for each energy bin beside the data points,
as well as for all energies in the legend.
The assumed source flux is $dN/dE=2.9 \times 10^{-8} (E/{\rm TeV})^{-1.85} {\rm cm^{-2} s^{-1} TeV^{-1}}$ (solid line),
representing a power-law extrapolation of the {\it Fermi} LAT spectrum of GRB 080916C
scaled to $z=$6.5 and $t=$50 sec after trigger (similar to interval d).
}
\label{fig:GRB080916C_z6.5}
\end{figure}

\begin{figure}[!htb]
\vspace{-0.5cm}
\centerline{\hspace{0cm}\vspace{0cm}\includegraphics[width=1\columnwidth]{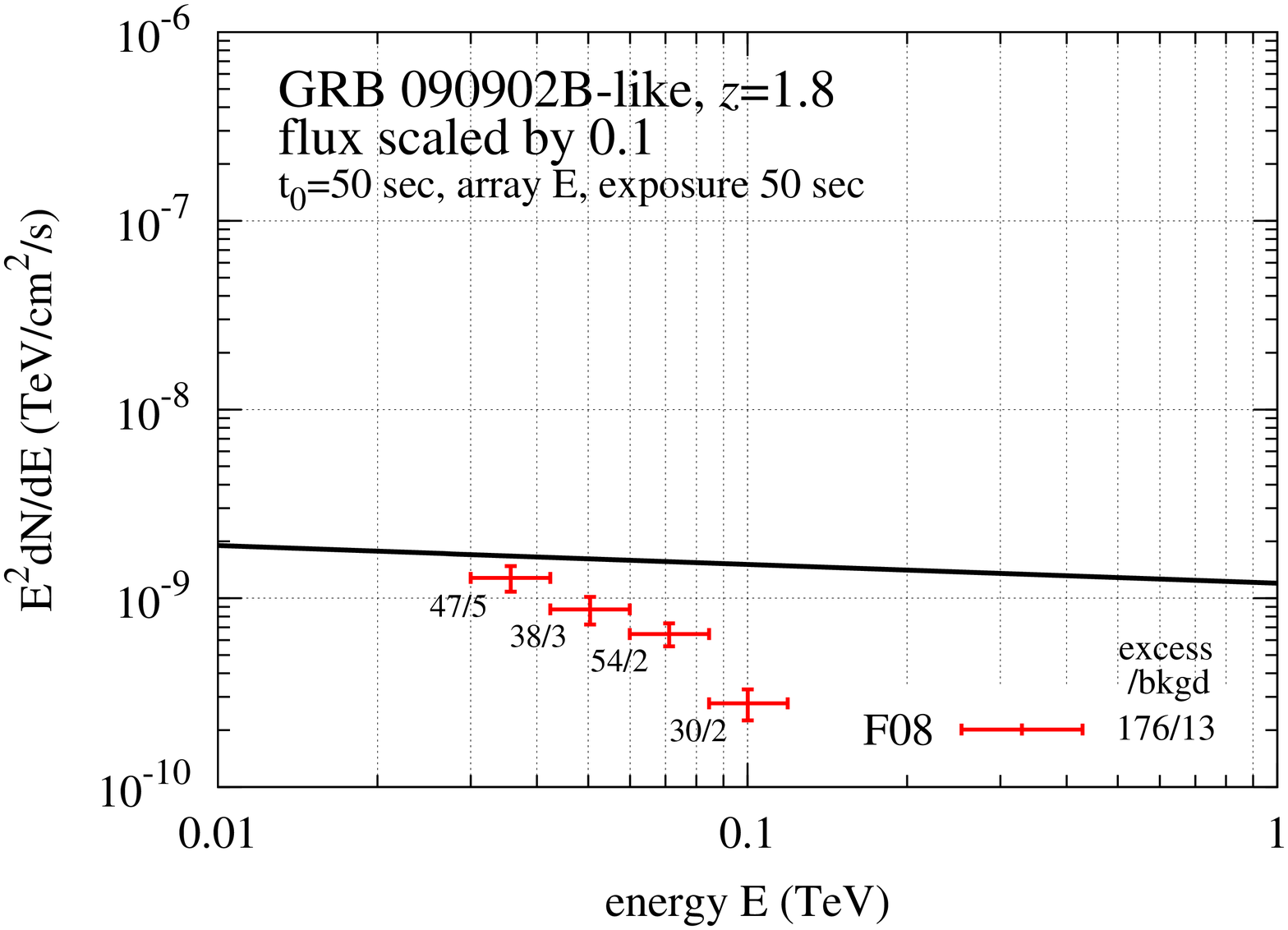}}
\centerline{\hspace{0cm}\vspace{0cm}\includegraphics[width=1\columnwidth]{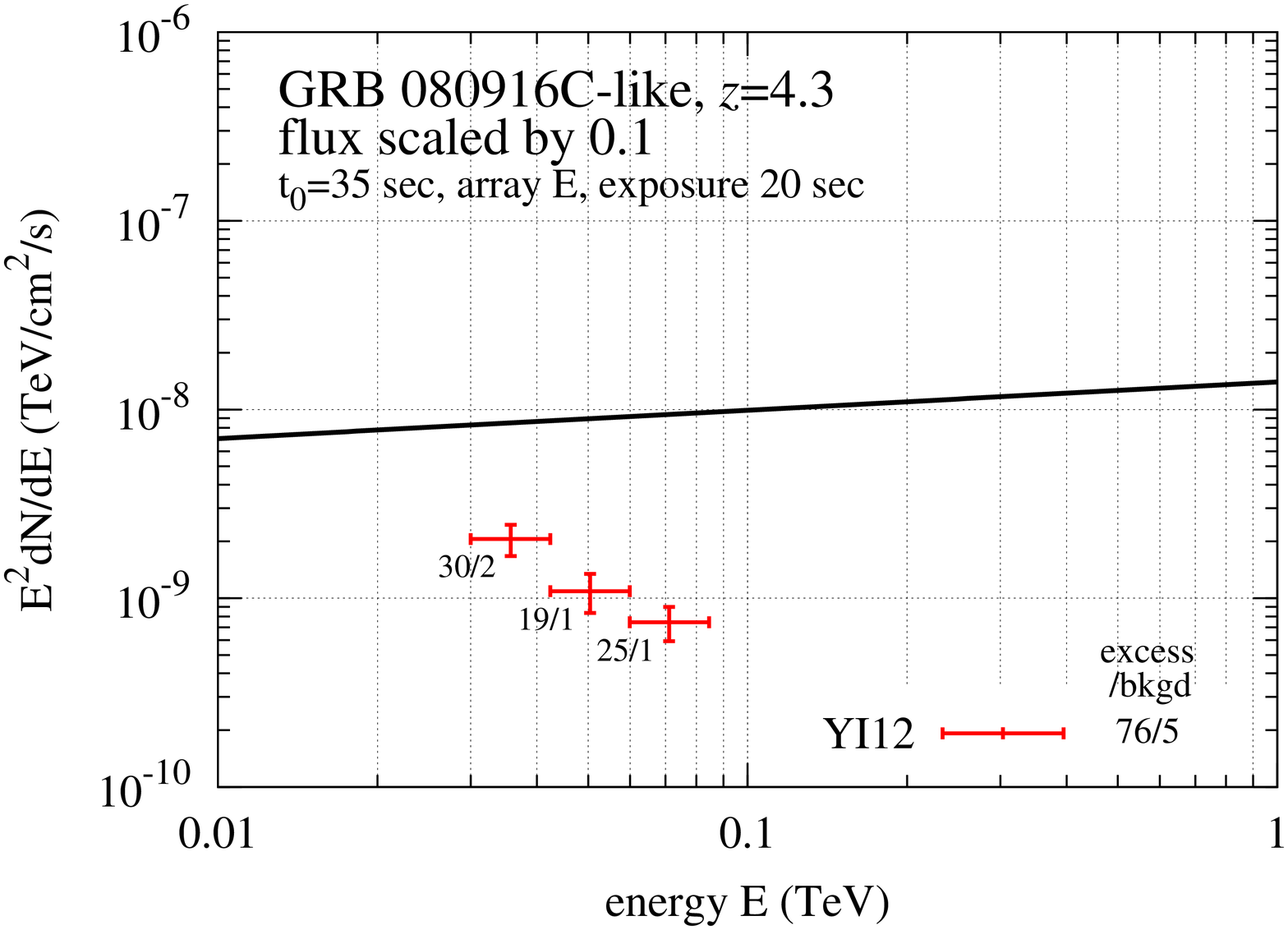}}
\caption{
Top:
Simulated spectrum for a GRB 090902B-like event at $z=$1.8 but with flux scaled by factor 1/10,
for exposure time 50 sec and array configuration E, adopting the EBL model of Franceschini et al \cite{Franceschini08}.
The number of detected photons/background events are denoted for each energy bin beside the data points,
as well as for all energies in the legend.
The assumed source flux is $dN/dE=1.2 \times 10^{-9} (E/{\rm TeV})^{-2.1} {\rm cm^{-2} s^{-1} TeV^{-1}}$ (black solid line).
Bottom:
Simulated spectrum for a GRB 080916C-like event at $z=4.3$ but with flux scaled by factor 1/10,
for exposure time 50 sec and array configuration E, adopting the EBL model of Y. Inoue et al \cite{YInoue12}.
The number of detected photons/background events are denoted for each energy bin beside the data points,
as well as for all energies in the legend.
The assumed source flux is $dN/dE=1.4 \times 10^{-8} (E/{\rm TeV})^{-1.85} {\rm cm^{-2} s^{-1} TeV^{-1}}$ (black solid line).
}
\label{fig:GRB_flux01}
\end{figure}

Some selected results are displayed in Figs.\ref{fig:GRB090902B_z1.8}-\ref{fig:GRB_flux01}.

For GRB 090902B, LAT detected 1 photon above 30 GeV at $t_0 \sim$80 sec,
while the CTA simulations for 50 sec exposure near this $t_0$ result in $\sim$1000-2000 photons depending on the EBL
(Fig.\ref{fig:GRB090902B_z1.8}).
This is roughly consistent with the expected factor $\sim 10^4$ difference in effective area at 30 GeV \cite{Funk12}.

Luminous bursts at low $z$ can result in enormous numbers of detected photons (Fig.\ref{fig:GRB090902B_z1.0})
and permit detailed studies of light curves and time-resolved spectra (Section \ref{sec:simulation-lc}),
from which we may delve into many issues left unsolved by {\it Fermi}.
For $z=1$, note also the potentially significant detection even up to $\sim$400 GeV
and after EBL attenuation by $\sim$3 orders of magnitude.
However, this postulates that the whole array has been slewed sufficiently rapidly, whereas in reality,
the sensitivity at the higher energies from the MSTs/SSTs may not be available, at least not very rapidly.
Such simulations and comparison with those for different subarray combinations
should be useful for addressing the relative merits/demerits of slewing the non-LST components.

Distant GRBs may also be detectable,
which would serve as powerful probes of the EBL at $z>2$,
beyond the expected reach of AGNs (Fig.\ref{fig:GRB080916C_z4.3}),
and possibly even into the cosmic reionization epoch at $z>6$ (Fig.\ref{fig:GRB080916C_z6.5}).
Even for some EBL models predicting significant absorption down to $\sim$10 GeV \cite{SInoue10}
with seemingly no hope for CTA, the simulations surprisingly reveal that a detection may be possible,
notwithstanding heavy attenuation.
This is needless to say for less opaque EBL models \cite{Gilmore12,YInoue12}.

However, as discussed in detail in Section \ref{sec:rates},
because of the low duty cycle and zenith angle constraints inherent in IACT observations,
the probability for CTA to detect bursts with such high luminosities and/or very high redshifts may be limited.
In this regard, we chose to simulate more common events with moderate luminosities in a simple way
by scaling the fluxes of our template cases by a factor of 1/10 (Fig.\ref{fig:GRB_flux01}).
This still leads to detection of up to a few hundred photons per burst,
certainly allowing valuable studies of the various science goals discussed in Section \ref{sec:science-case}.
Even non-detections of bursts at $z>2$ can provide important new constraints on the high-$z$ EBL.

One rather alarming fact is that for most cases shown here, EBL absorption
is affecting almost the entire energy range over which photons are detected,
even for the lower redshift bursts.
This can prevent us from reliably ascertaining the intrinsic spectrum before EBL absorption
and leave large uncertainties in the resulting constraints,
unless simultaneous measurements with {\it Fermi} LAT can be performed (see Section \ref{sec:MAGIC}).

Fortunately, for all cases, in the lowest plotted energy bin of 30-40 GeV,
the number of excess over background events is still significant.
This gives us hope that the detection energy threshold can be lowered appreciably for GRBs
by going beyond standard analysis criteria as implemented in D. Mazin's tool
and allow access to the spectral region unaffected by EBL absorption.
Array B seems to be the best for this goal, but array E does not appear significantly worse;
when compared, the latter results in about 30-50 \% less photons,
but may manage to do a qualitatively competitive job in determining spectra, for either bright or moderate events.
Yet the real strength of the B-type configurations may be achieving the lowest thresholds possible,
which cannot be properly appreciated with the current tool, and warrants deeper examination.


\subsection{Simulations of light curves}
\label{sec:simulation-lc}

As a first demonstrative study,
we have also carried out simulations of GRB light curve measurements,
assuming the following:
\begin{itemize}

\item Take as template GRB 080916C.

\item Normalize the time-dependent flux with the light curve at $E>0.1$ GeV as measured by {\it Fermi} LAT.

\item Extrapolate the spectra for specific time intervals
to higher energies using the spectral indices as determined by LAT for each interval.

\item Adopt the EBL model of \cite{Razzaque09}.

\item Consider array configuration E and zenith angle of observation 20 deg.

\item Taking a given time bin as the exposure time, evaluate the flux in a given energy band
that would be measured by CTA with the aforementioned simulation tool.
Repeat the procedure for all time bins to produce a light curve.

\item Consider also a burst with the same properties
but with fluxes scaled by factor 1/10
so as to simulate events with more moderate luminosities.

\end{itemize}

\begin{figure}[!htb]
\centerline{\includegraphics[width=0.95\columnwidth]{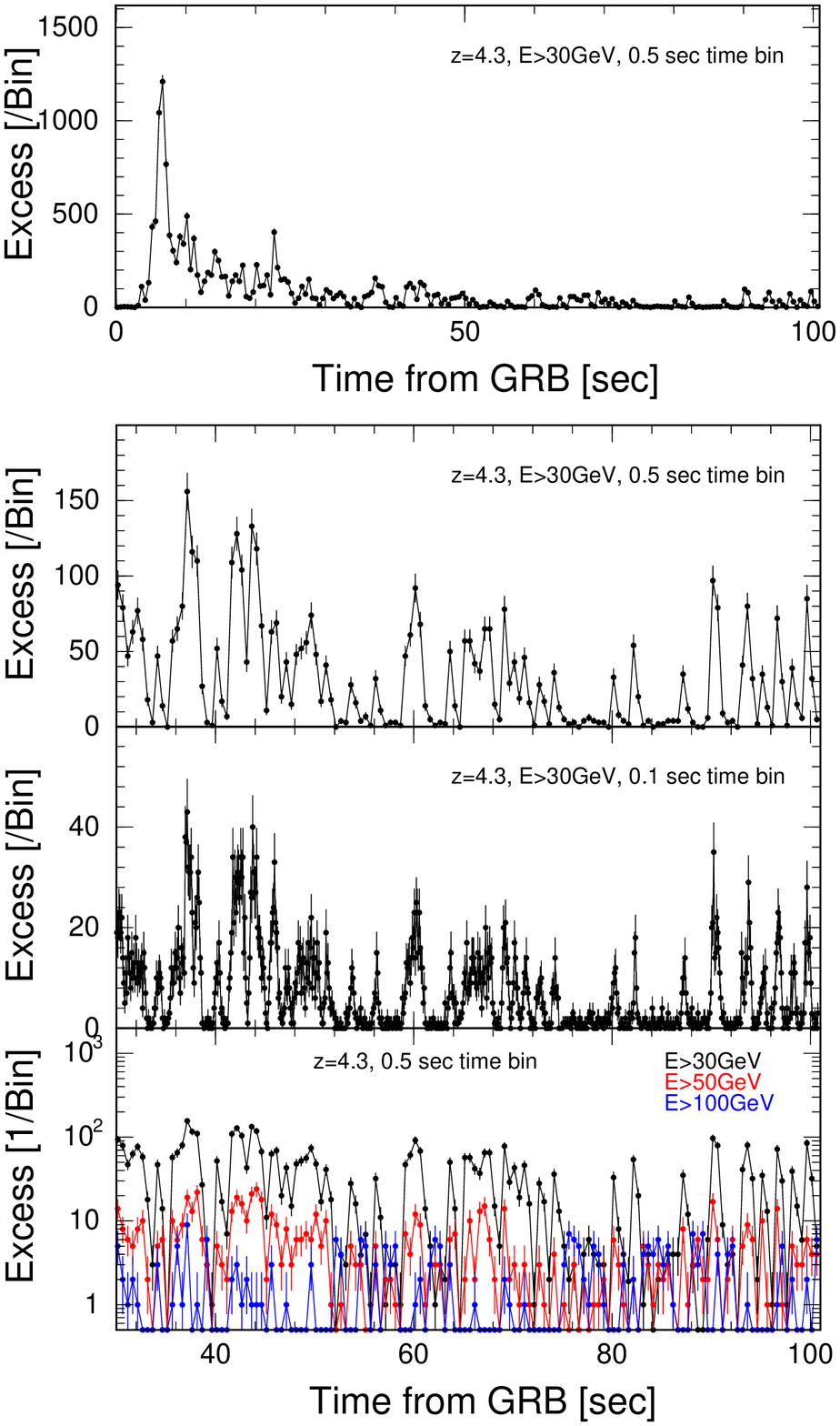}}
\caption{Simulated light curves of GRB 080916C at $z=4.3$ for CTA array E.
The EBL model of \cite{Razzaque09} was assumed.
Top: Light curve for $E>30$ GeV from $t_0=0$ sec, with 0.5 sec time binning.
Upper middle: Same as top panel, but plotted from $t_0=30$ sec.
Lower middle: Same as upper middle panel, but with 0.1 sec time binning.
Bottom: Light curves from $t_0=30$sec with 0.5 sec time binning,
for $E>30$ GeV, $E>50$ GeV and $E>100$ GeV, from top to bottom.
}
\label{fig:lcsim_GRB080916C}
\end{figure}

\begin{figure}[!htb]
\centerline{\includegraphics[width=0.95\columnwidth]{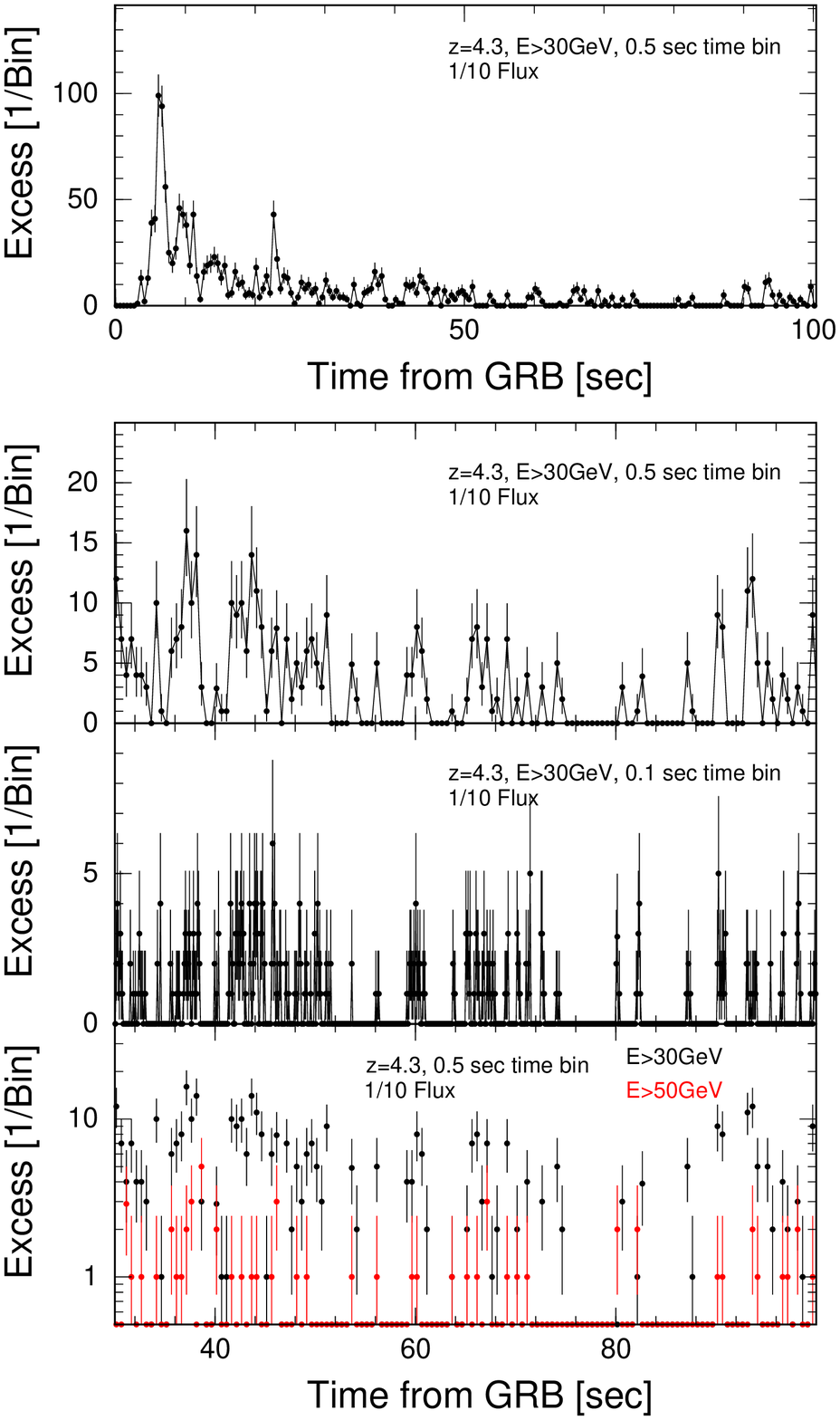}}
\caption{Same as Fig.\ref{fig:lcsim_GRB080916C}, but with assumed flux scaled by factor 1/10.}
\label{fig:lcsim_GRB080916C_flux01}
\end{figure}

The results are shown in Figs.\ref{fig:lcsim_GRB080916C}-\ref{fig:lcsim_GRB080916C_flux01} .
Thanks to its significantly larger effective area compared to {\it Fermi},
CTA is potentially capable of resolving the light curve in exquisite detail
for such bright bursts (and to a lesser extent for moderately bright bursts),
as long as it can begin observing during the prompt phase.
The energy-dependent light curves
\footnote{Note that for the higher energy bands, some time bins with too few photons
may not allow reliable flux measurements, so that studies of the energy-dependent variability
will be restricted to the brighter pulses in the light curve.}
would be especially valuable
for extracting crucial information on the physics of the emission mechanism (Sections \ref{sec:GRBphys}-\ref{sec:UHECR}).
In particular, it could reveal definitive signatures of hadronic emission processes (see Fig.\ref{fig:MeVGeVLC})
that was impossible with time-integrated spectra alone.
It will also be valuable for distinguishing whether spectral cutoffs
are due to EBL attenuation or physics intrinsic to the GRB (Section \ref{sec:EBL}; see also \cite{Mazin12}),
not to mention searches for Lorentz invariance violation (Section \ref{sec:LIV}).
These aspects will be investigated more quantitatively in the future
by incorporating the relevant physics in more detail.
For all these goals, an energy threshold as low as possible is strongly desirable
in order to achieve the broadest spectral lever arm
over which we can exploit the energy-dependent variability.
Corresponding studies for the afterglow emission are also forthcoming.

\section{Detection Rate Expectations}
\label{sec:rates}

We now discuss expectations for the detection rate of GRBs with CTA.
Two independent approaches are presented,
one by Gilmore et al. (see also \cite{Texas2010, CTAGRB}) and another by Kakuwa et al. (see also \cite{Kakuwa12}).
Although they share some similarities in the assumptions,
the main difference lies in the modelling of the GRB population,
the former based directly on observed GRB samples,
and the latter using a somewhat more theoretical method.
The treatment of the CTA performance is also different;
Gilmore et al. employ a phenomenological model,
whereas Kakuwa et al. utilizes the official CTA performance files.
The results obtained through the two approaches are generally within a factor of 2 of each other
and can be considered consistent.

Note that the ``delay time'' as used below refers to the sum of all types of delays
between the satellite-onboard burst trigger and the start of targeted CTA observations,
including the time for the GRB alert to reach the telescopes from the satellite,
any other kind of delay before the telescopes can start slewing, as well as
the slewing time of the telescopes.


\subsection{Observation-based population model}
\label{sec:rates-Gilmore}

Despite being high-priority targets for current IACTs,
GRBs have so far escaped detection at VHE and only yielded flux upper limits despite dozens of follow-up attempts
(Section \ref{sec:status-VHE}).
High hopes come with the CTA observatory to finally succeed in this endeavor,
thanks in particular to its order of magnitude improvement in sensitivity and much lower energy threshold.
Unfortunately, because of their transient nature, GRBs are unlikely to be observed serendipitously
in the limited field-of-view of IACTs.
For example, assuming a whole sky rate of $\sim600$ GRBs/year, a $5^{\circ}$ diameter FoV and a $10\%$ duty cycle,
the telescopes will cover a patch of the sky where a GRB is expected to go off only once every $\sim 35$ years.
(However, note that CTA will also have the ability to observe in a wide-field mode
by splitting its array of MSTs so as to increase its sky coverage at a given time,
albeit with a reduced sensitivity, a strategy already foreseen for survey and monitoring purposes;
see Section \ref{sec:widefield}).

In order to significantly increase the chances of detecting VHE emission from GRBs,
an essential strategy is to follow-up on external GRB alerts to point the telescope array toward a localized GRB.
Such GRB alerts are provided by various space-based missions through the Gamma-ray Coordinates Network (GCN).
Which GRB-dedicated missions will still be operating at the start of CTA operation is fairly uncertain
and depends mainly on the contingency of mechanical failures and funding issues.
Currently the following two instruments are providing the large majority of GCN alerts:

\begin{itemize}
\item {\it Swift} BAT \cite{Gehrels09}: alert rate $\sim 95$ GRBs/year with extremely good localization ($\lesssim 10$ arcmin)
\item $Fermi$ GBM \cite{Meegan09}: alert rate $\sim 250$ GRBs/year with poor localization (several degrees)
\end{itemize}

SVOM \cite{Schanne10,Paul11}, a French-Chinese mission dedicated to the study of GRBs,
is expected to be launched near CTA's first light.\footnote{Expected launch in $\gtrsim 2017.$}
It will provide GCN alerts with very similar characteristics as {\it Swift} alerts:
alert rate $\sim 70-90$ GRBs/year and excellent localization well within the CTA field-of-view.
Here we estimate the probability for CTA to detect VHE emission from these
two very different types of GCN alerts: {\it Swift}-like (i.e. from {\it Swift} itself or SVOM) and GBM-like.
The results presented in this section are taken from an in-depth analysis published in \cite{CTAGRB},
to be consulted for more details.

A Monte-Carlo simulation was used to model the performance of the CTA array
\footnote{Only the LST and MST array were considered, as the energy range of the SSTs (threshold $\sim 1$ TeV)
is not suitable for studying distant extragalactic sources on account of the expected spectral attenuation in the EBL.}
as well as the properties of the GRB emission at VHE derived from a phenomenological approach,
which uses temporal and spectral information of GRBs detected by {\it Fermi} LAT
and other instruments operating at sub-MeV energies ({\it Fermi} GBM, {\it Swift}, BATSE).
Because of our lack of knowledge on the spectral characteristics of GRB very high-energy emission,
we constructed two spectral models in between which we reasonably expect the true GRB population to lie:

\begin{itemize}
\item `Bandex' model: a simple extrapolation of the Band function
\footnote{Band function parameters are all drawn from BATSE distributions, although for {\it Swift} simulated bursts,
a global fluence multiplier of 0.75 is applied to the BATSE fluence distribution to provide the best fit between
BATSE and {\it Swift} BAT fluences in the $15-150$ keV range.}
to VHE with a maximum limit of -2.0 for the high energy index.

\item `Fixed' model: a power-law component is added to the Band function, with a fixed spectral index of -2.0
and a normalization chosen so that the energy flux ratio between the LAT ($100 \mbox{ MeV}-300\mbox{ GeV}$)
and BATSE ($50-300 \mbox{ keV}$) energy range is 10\%.
\end{itemize}

We note that these two spectral scenarios are consistent with the LAT detection rate of $\sim 10$ GRBs/year
and more specifically the `Bandex' and `Fixed' models are consistent with the spectral behavior of
GRB 080916C \cite{080916C} and GRB 090902B \cite{090902B} respectively.
The redshift distribution of observed GRBs was derived from $\sim 170$ {\it Swift}-detected GRBs
and all spectra were naturally convolved with the EBL model of \cite{Gilmore12}.
However, we stress that intrinsic spectral curvature is not considered in our model
mostly because of the large theoretical uncertainty of this feature.
As a consequence of this caveat, the actual CTA detection rate could be lower
than our predictions if strong intrinsic curvature below $\sim 100$ GeV is common in GRB spectra.
Substantially higher detection rates are considered rather unlikely. 
The VHE light curve was assumed flat during the prompt emission to which a duration $T_{90}$
(drawn from the $T_{90}$ BATSE distribution) was assigned.
Extended VHE emission was modeled assuming a temporal decay similar to the one measured
for bright LAT GRBs \cite{Ghisellini10}, proportional to $(t/T_{90})^{-1.5}$.

In order to investigate the chances of detecting the prompt emission, we have also considered
the case of light curves where no high energy emission emerges after the prompt phase,
which are labelled `prompt'.

The performance of the LST and MST arrays were derived independently
from a simple interpolation of the known performance of current IACTs
(more specifically, the VERITAS effective area and background rate).
The VERITAS energy threshold ($\sim 100 $ GeV) was shifted toward lower energies
by the ratio of the primary mirror collecting areas ($\sim 4$ for LSTs and $\sim 1$ for MSTs).
The normalization was then increased assuming a linear scaling
with the number of telescopes (assuming 4 LSTs and 25 MSTs).
Finally, the background rate was extrapolated to these new effective areas assuming a spectrum $E^{-2.7}$.
To allow for uncertainties in the performance of the LST and MST arrays,
we also simulated an `optimistic' CTA performance.
For this, we further reduced the LST low-energy threshold from 25 GeV to 10 GeV
(which might be achieved with an improved trigger system),
increased the MST effective area by a factor of 3 (to consider additional MST telescopes up to $\sim 75$)
and decreased the background rate for both the LST and MST arrays by a factor of 3
(to consider improved performance through event containment regime and advanced analysis techniques).
Lastly, we assumed a typical 60 sec (100 sec) delay for the LSTs (MSTs) to point toward a localized GRB.
\footnote{This includes the delay for GCN alerts to be sent out to the ground,
the response time of the observers as well as the slewing time for the telescopes to be on target.}
Although this is shorter than the typical values realized in current IACT observations (e.g. \cite{Garczarczyk09}),
we allow for the possibility that future improvements to the GCN and telescope alert procedures
and observer response time could lower the delay times.

For a GRB to be detected by CTA, a first necessary criterion is that it is observable by the array at the time of the alert.
In our study, we assume a 10\% duty cycle although we recognize that observations under moonlight
(albeit with a higher energy threshold) could increase the duty cycle to $\sim 13\%$ or more. 
We considered CTA capable of observing a GRB when its zenith angle is smaller than $75^{\circ}$
and we parametrized the increase in energy threshold with zenith angle as:
$E_{th}({\rm Zenith}) = E_{th}(0) \times \cos({\rm Zenith})^{-3}$.
Finally, we also included the effect of the anti-solar bias present in {\it Swift}-detected GRBs,
as discussed in \cite{GilmoreGRB}.

In case the simulated GRB was deemed observable, the significance for various observation timescales
was computed following the procedure described in \cite{LiMa}.
Depending on the phenomenological emission model
and array performance used,
the average detection rates for {\it Swift}-like alerts obtained
for one array site (either CTA North or South only) are summarized in Table~\ref{table:GRBrate_obs}.

\begin{table*}[htb]
\caption{
Expected detection rates for {\it Swift}-like alerts in the observation-based model by Gilmore et al., for one array site.
The numbers should be doubled when considering both CTA North and South.
}
\begin{center}
\begin{tabular}{lcccc}
                                    & Bandex & Fixed & Bandex prompt & Fixed prompt \\ \hline
    baseline [yr$^{-1}$] & 0.35 & 0.6 & 0.13 & 0.21 \\
    optimistic [yr$^{-1}$] & 0.8 & 1.6 & 0.28 & 0.54
\end{tabular}
\end{center}
\label{table:GRBrate_obs}
\end{table*}

In comparison, the estimated detection rates are around $\sim 0.1-0.2$ GRB/year for current IACTs.
These numbers would naturally be increased by a factor of $\sim 2$
as long as both CTA North and South are built with similar numbers of LSTs.
In case of detection, CTA will provide photon statistics about an order of magnitude
higher than achievable with currently operating IACTs.
In particular, CTA-detected bursts will have significant statistics below $100$ GeV
where the spectrum is hardly absorbed by the EBL, which makes such detection a golden case for EBL studies.

\begin{figure}[!htb]
\centering
$\begin{array}{cc}
\includegraphics[width=2.9in,height=2.6in]{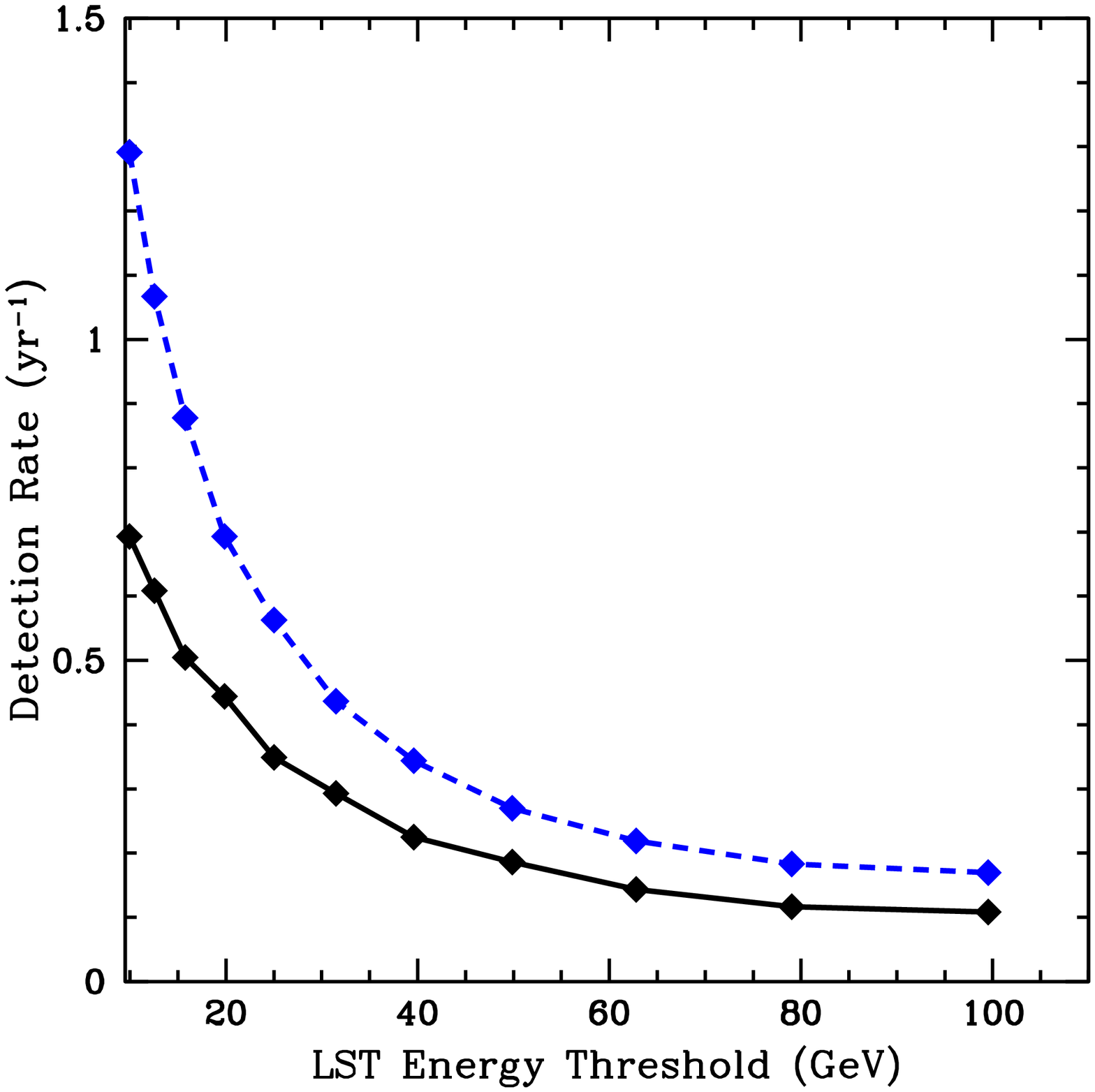}\\
\includegraphics[width=2.9in,height=2.6in]{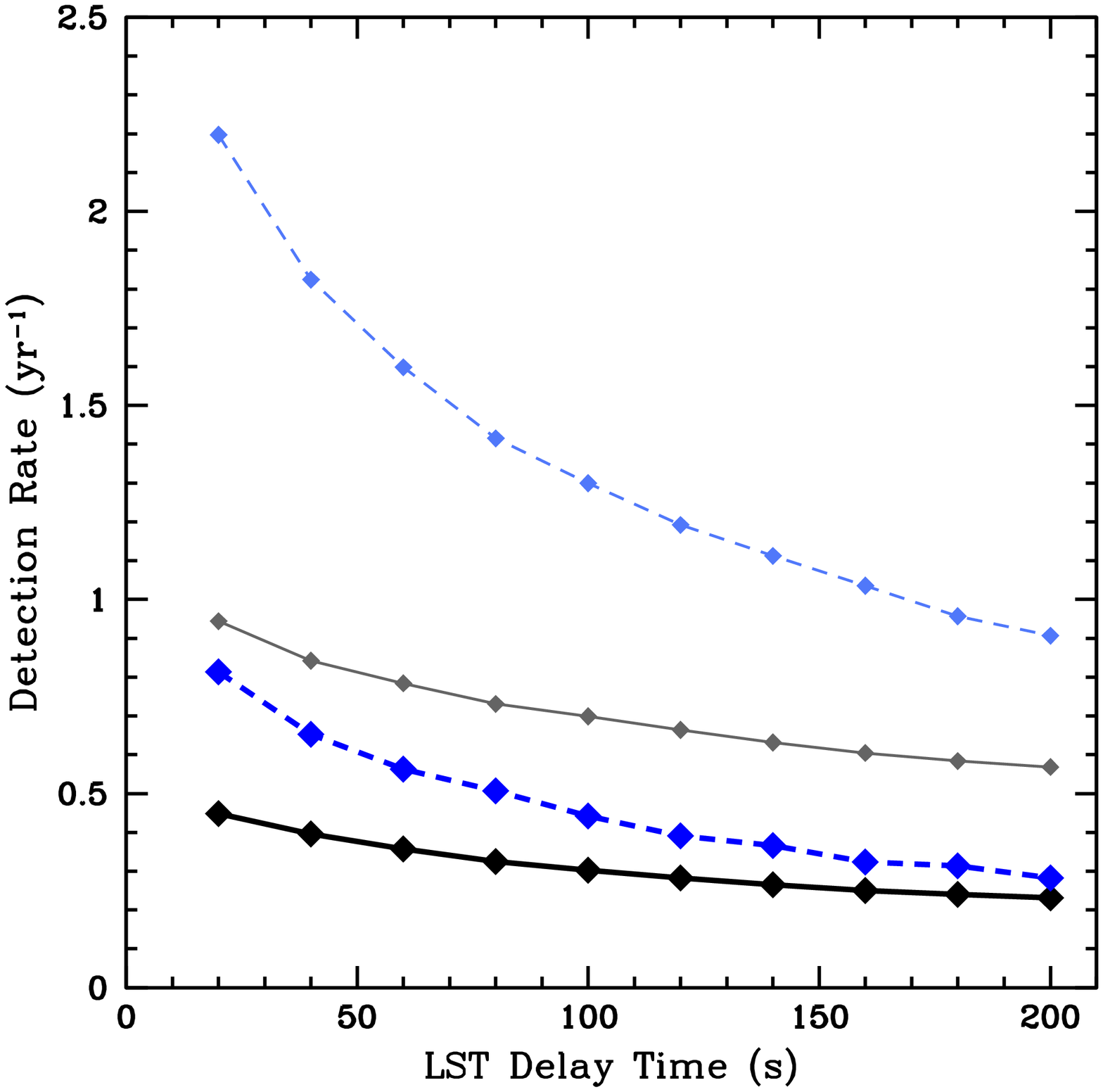}
\end{array}$
\caption{{\bf Top:}
GRB detection rate for one CTA site and {\it Swift}-like alerts as a function of the LST energy threshold.
Solid black: `bandex' model; dashed blue: `fixed' model.
Background rate from the baseline CTA performance is assumed.
{\bf Bottom:}
GRB detection rate for one CTA site and {\it Swift}-like alerts as a function of the LST delay time.
Solid dark black: `bandex'/baseline; Dashed dark blue: `fixed'/baseline;
Solid grey: `bandex'/optimistic; Dashed light blue: `fixed'/optimistic.
The LST delay time includes the delay for GCN alerts to be sent to the ground
and the response time of the observers as well as the slewing time for the LSTs to arrive on target.
The standard delay time used in our simulations is 60s.
The numbers should be doubled when considering both CTA North and South.
}
\label{DR_Eth}
\end{figure}

We also investigated how the average GRB detection rate varies as a function of critical instrument parameters.
The low-energy threshold was found to be the most important for CTA's capability to detect GRBs.
For example, if the energy threshold were decreased from $\sim 25$ GeV to $\sim 10$ GeV,
the average detection rate would increase by a factor of $\sim 2$ (see Fig. \ref{DR_Eth}, top panel).
It is therefore crucial to develop an efficient triggering scheme in order to lower the energy threshold of the LSTs
as much as possible.
\footnote{Building the array at higher altitude would also be beneficial.}
The time delay for the telescopes to slew onto the burst position is an important parameter
if one wants to acquire photon statistics early during the burst light curve and hopefully catch
the end of the prompt emission of long GRBs.
However, changing the time delay from $\sim 40$s to $\sim 80$s changes the burst detection rate
by only $\sim 20\%$ (see Fig. \ref{DR_Eth}, bottom panel).
The LST design already has $\sim 20$s slewing time to any position in the sky so that it seems difficult
to further increase the number of GRBs with this parameter.
GRB science may benefit from minimizing other sources of time delay, such as GCN reporting time
or that due to operator intervention upon receiving a GRB alert.
Finally, as previously mentioned, a $\sim 30\%$ or more increase in observing time could be obtained
by performing moonlight observations (although with a higher energy threshold),
an improvement in duty cycle which seems very reasonable to reach.

Finally, we investigated the case of GBM-like alerts which come at a much higher rate but with quite a poor localization
\footnote{We also took into account the fact that GBM bursts are significantly brighter than {\it Swift} bursts,
on average by a factor of $\sim 3$.}.
The statistical plus systematic error radius ranges from $\sim 3^{\circ}$ to $\sim 20^{\circ}$
which is in many cases larger than the LST field-of-view ($\sim 4.6^{\circ}$ diameter).
\footnote{We point out that reducing GBM localization uncertainties is a key objective of the GBM team
which might well succeed in further reducing GBM error radius by the time of CTA's first light.}.
A simple strategy where the whole CTA array is pointed to the best burst localization yields very poor results
(detection rate $\sim 10$ times lower than for {\it Swift}-like alerts).
However, it is possible to implement smarter observing strategies,
such as a `scanning mode' where the LSTs would scan over the whole GBM error box \cite{OrbitMode},
a `divergent pointing mode' where the LSTs would be initially offset so as to cover the entire error box,
or an `array splitting mode' where the MST array would be divided into sub-arrays to cover a larger part of the sky
(although the latter two strategies would have the disadvantage of increasing the energy threshold of the observation).
We found the scanning mode to yield a detection rate about half that for {\it Swift}-like GRBs,
which makes GBM alerts competitive in terms of detection rate
although photon statistics will certainly suffer from such strategies.


\subsection{Theory-based population model}
\label{sec:rates-Kakuwa}

Using a somewhat more theoretical approach for modelling of the GRB population than in the preceding section,
here we independently estimate the detection rate with CTA
and study how the results depend on properties such as array configuration, delay time, etc.
For this section, we mainly consider {\it Fermi} GBM as the GRB alert facility,
although estimates for the case of SVOM alerts are also given at the end.
These can be compared with the respective results for GBM-like and {\it Swift}-like alerts of the previous section.
More details together with complementary discussions are presented in a separate publication \cite{Kakuwa12}.
We adopt the cosmological parameters
$H_0 = 70$~km~s$^{-1}$Mpc$^{-1}$, $\Omega_{m} = 0.3$ and $\Omega_\Lambda = 0.7$.

\noindent{\bf Intrinsic GRB Properties.}
We simulate the GRB population using Monte Carlo methods.
Following \cite{Wanderman10},
the GRB luminosity function, i.e.
the GRB rate per unit comoving volume at redshift $z$
per logarithmic interval of 1--$10^4$~keV peak luminosity $L_p$
is assumed to be
$\Psi(L_p,z) = \rho(z)\phi(L_p)$,
where 
$\rho(z)\propto(1+z)^{2.1}$ for $z<3.1$,
$\rho(z)\propto(1+z)^{-1.4}$ for $z>3.1$,
$\phi(L_p)\propto L_p^{-0.17}$ for $L_p<10^{52.5} {\rm erg~s^{-1}}$, and
$\phi(L_p)\propto L_p^{-1.44}$ for $L_p>10^{52.5} {\rm erg~s^{-1}}$,
which is consistent with the observed characteristics of {\it Swift} GRBs.
From the measured properties of a sample of GBM bursts,
we also determine the relations between $L_p$, time-averaged luminosity $L_a$
and isotropic-equivalent energy ${E}_{\rm iso}$ as
$L_a= 0.31L_p$ and 
$\log E_{\rm iso,52}=0.9\log L_{p,52}+0.6$.
Then the duration evaluated by
$T_{90}=(1+z) E_{\rm iso}/L_a$ 
forms a distribution that provides a good match
to the one observed by GBM for long GRBs.

The light curve of the prompt emission is assumed to have a top-hat shape
with luminosity $L_a$ and duration $T_{90}$.
Its fiducial spectrum is described by the Band function \cite{Band93},
with normalization fixed by $L_a$ as above,
the spectral peak energy $E_p$ given by the observed $E_p$--$L_p$ relation \cite{Yonetoku04, Ghirlanda09},
and the low-energy and high-energy spectral indices $\alpha$ and $\beta$
sampled from the distribution observed in bright BATSE bursts \cite{Kaneko06} with the restriction $\beta<-2$.
We also consider cases with an extra spectral component with photon index -2
and whose 0.1--100~GeV luminosity is a fraction $R_{\rm extra}=0.1$ of $L_a$. 
The high-energy afterglow is characterized by a spectrum with fiducial photon index $p_E=-2$
and 0.1--100~GeV luminosity that evolves as
$L_{\rm AG}(t) = 10^{52} {\rm erg}~{\rm s}^{-1}
(E_{\rm iso}/10^{54}{\rm erg})
(t/(1+z)10~{\rm s})^{p_t}$
as a function of postburst observer time $t$,
taken to be nonzero only for $t>T_{90}$
and with fiducial temporal decay index $p_t=-1.5$ \cite{Ghisellini10}.

We adopt the EBL model of Razzaque et al. \cite{Razzaque09} which is limited to $z<5$,
so the redshift range $z>5$ cannot be treated in this calculation.
For comparison, the EBL model of Kneiske et al. \cite{Kneiske04} is also employed.

\noindent{\bf CTA Follow-up Observations.}
For GBM alerts,
the trigger threshold in peak photon flux is taken to be
$1.5\;{\rm ph}\;{\rm cm}^{-2}\;{\rm s}^{-1}$ in the $8 - 10^3\;$keV band,
which is satisfied by $90\%$ of actual GBM bursts.
Follow-up with CTA will be feasible for only a fraction of them
that is sufficiently well localized so that they can be reasonably covered by the FoV of the LSTs.
Here we choose the criterion for initiating follow-up to be when the GBM error radius is $< 5\;$deg
(note the current condition of $< 4\;$deg for MAGIC; Section \ref{sec:MAGIC}).
Compared with the $\sim 4.6\;$deg diameter currently foreseen for the LST FoV,
this implies that a considerable fraction of the bursts can be missed by falling outside the FoV.
Although the actual situation would vary somewhat from burst to burst,
we approximate the probability that such GBM bursts are still caught within the LST FoV
with a constant value of $\sim 0.1$ (see \cite{Kakuwa12} for more details),
which is incorporated in all calculations below.
Nontrivial LST follow-up strategies
such as divergent initial pointing (currently under study by the CTA Monte Carlo simulation group)
or scanning of the GBM error circle \cite{OrbitMode},
as well as future improvements in the GBM localization algorithm 
can significantly increase this probability,
the quantitative effects of which will be discussed in subsequent studies.
We also evaluate the probability that a given GBM localization accuracy is realized as a function of the fluence
by making use of actually measured values as reported in the GCN.
The delay time $T_{\rm delay}$ between the burst trigger and the start of CTA observations
is assumed to obey a log-normal distribution,
with a fiducial peak at $\tau_{delay} = 100\;$s, dispersion $\sigma_{\rm delay} = 0.4\;$dex,
and a lower bound of $T_{\rm delay}>20\;$s.
This accounts for a plausible degree of improvement from the delay times
actually realized during MAGIC-I observations in 2005-2008,
which can be fit by a similar distribution but with $\tau_{delay} = 160\;$s and $\sigma_{\rm delay} = 0.5\;$dex,
and for which the average telescope slewing time was $\sim 90\;$s
(c.f. \cite{Albert07,Garczarczyk09}).

For the performance of CTA, we make use of the information provided by the CTA Monte Carlo simulation group
such as effective area, background rate and energy resolution,
particularly from the simulations of the Heidelberg group \cite{Bernloehr12}.
The array configurations B, D, and I, are considered, the latter being the fiducial choice.
Since the performance files are currently available only for zenith angles $\theta_{zen} = 20$ and $50\;$deg,
the results are presented for both of these values,
i.e. supposing that all GRBs are observed either at $\theta_{zen} = 20$ or $50\;$deg;
the true situation should lie in between.
We also employ the performance files for which the event selection has been optimized
for $0.5\;$hour exposure time, the shortest available at the moment.
For the minute-timescale exposures more appropriate for GRB observations,
the actual performance is expected to be better
as the background rate should be lower and the effective area higher \cite{Funk12}.
The duty cycle is taken to be 10~\%,
although it may be increased up to $\sim$15~\% under moonlight
at the expense of a higher energy threshold and lower sensitivity.
We also set a limit of $\theta_{\rm zen}<60$~deg on the observable range of zenith angles.
\footnote{The contribution from $\theta_{\rm zen}>60$~deg is expected to be small
on account of the correspondingly high energy threshold.}

In accord with the standard criteria for IACT observations \cite{Aharonian01,CTA11},
a detection is declared
when the number of photons $N_{\gamma,i}$ obtained
within a given energy interval during a given exposure time satisfy all of the following:
(1) $N_{\gamma,i} > N_{\rm min}$,
(2) $N_{\gamma,i} > m\sqrt{N_{{\rm BG},i}}$, 
and (3) $N_{\gamma,i} > e N_{{\rm BG},i}$, 
where $N_{{\rm BG},i}$ is the number of background events,
$N_{\rm min} = 10$, $m=5$ and $e=0.05$.
We designate detection in terms of the differential flux
when the above applies to 
one or more energy bins whose widths correspond to the energy resolution.
The assumed exposure time is $T_{90} - T_{\rm delay}$ for the prompt emission
and up to 4~hours maximum for the afterglow.

\noindent{\bf Results.}
For convenience, we classify our model GRB sample into the following subsets:
\begin{itemize}
\item {\it Init}: GRBs triggered by GBM that satisfy $T_{90}>2$~sec and $z<5$.
\item {\it CTAobs}: GRBs belonging to {\it Init} whose $\theta_{\rm zen}<60$~deg,
localization accuracy $<5$~deg and occur during the 10 \% duty cycle,
i. e. bursts which are observable (but not necessarily detectable) by CTA.
\item {\it Pobs}: GRBs belonging to {\it CTAobs} that satisfy $T_{\rm delay}<T_{90}$,
i. e. bursts which are observable (but not necessarily detectable) during the prompt phase.
\item {\it Pdet}: GRBs that are caught within the LST FoV and whose prompt photons are detected.
\item {\it Adet}: GRBs that are caught within the LST FoV and whose afterglow photons are detected.
\end{itemize}

Table~\ref{table:GRBrate_theory1} summarizes the results for our fiducial parameters
$R_{\rm extra}=0$, $p_E=-2$, $p_t=-1.5$, $\tau_{delay} = 100\;$s and $\sigma_{\rm delay} = 0.4\;$dex,
for one CTA site.
The number for both CTA North and South will simply be double these values.
Note that the dispersion in the detection rates corresponds to the range of $\theta_{\rm zen}=$20--50~deg.
\begin{table*}[htb]
\caption{
Expected detection rates for the fiducial parameters of the model by Kakuwa et al., for one CTA site.
The numbers (except {\it Init}) should be doubled when considering both CTA North and South.
}
\begin{center}
\begin{tabular}{lcccccc}
    & {\it Init} & {\it CTAobs} & {\it Pobs} & {\it Pdet} & {\it Adet} \\ \hline
    {\it Fermi} GBM [yr$^{-1}$] & 200 & 1.8 & 0.66 & 0.013--0.033 & 0.09--0.2 \\
    SVOM ECLAIRS [yr$^{-1}$] & 56 & 2.0 & 0.65 & 0.09 & 0.53
\end{tabular}
\end{center}
\label{table:GRBrate_theory1}
\end{table*}

\begin{figure}[!htb]
\centerline{\includegraphics[width=0.9\columnwidth]{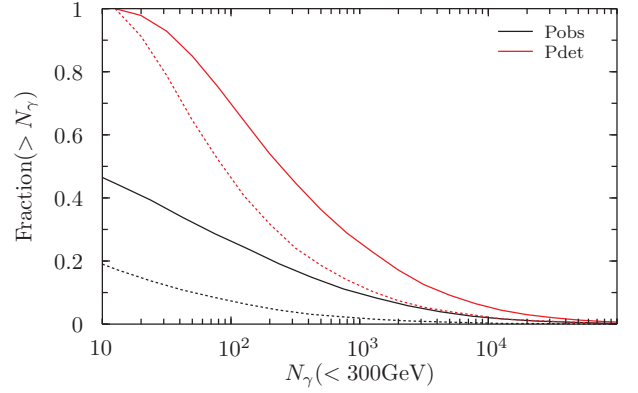}}
\caption{
Cumulative distribution of photon counts with energies $< 300$~GeV
in the case of prompt detections {\it Pdet} for our fiducial parameters (red curves),
compared with that for prompt observable events {\it Pobs} (black curves).
Solid and dashed curves correspond to $\theta_{\rm zen}=20$~deg and $\theta_{\rm zen}=50$~deg, respectively.
}
\label{fig:Ngam_fid_integral}
\end{figure}

Fig.~\ref{fig:Ngam_fid_integral}
shows the cumulative distribution of photon counts with energies below 300~GeV 
in the case of prompt detections,
compared with that for {\it Pobs}, events observable during the prompt phase by CTA.
When successfully detected,
we expect $N_{\gamma}>60$--150 with 60~\% probability
and $N_{\gamma}>450$ with 20~\% probability even for $\theta_{\rm zen}=50$~deg. 

\begin{figure}[!htb]
\centerline{\includegraphics[width=0.9\columnwidth]{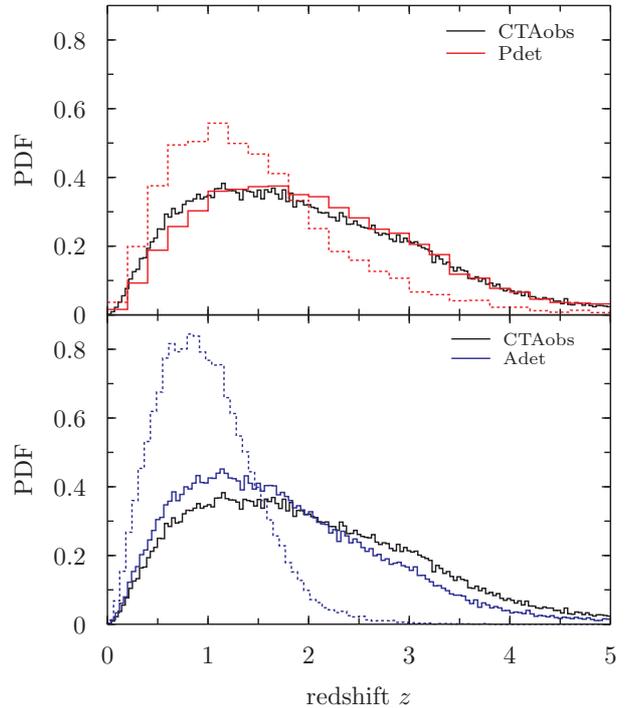}}
\caption{
Redshift probability distribution functions for our fiducial parameters
in the cases of prompt detections {\it Pdet} (red curves, top panel), afterglow detections {\it Adet} (blue curves, bottom panel),
and CTA observable events {\it CTAobs} (black curves).
Solid and dashed curves correspond to $\theta_{\rm zen}=20$~deg and $\theta_{\rm zen}=50$~deg, respectively.
}
\label{fig:z_fid_integral}
\end{figure}

In Fig.~\ref{fig:z_fid_integral},
we plot the redshift probability distribution functions (PDFs)
separately for the cases of prompt and afterglow detections,
compared with that for {\it CTAobs}, events observable by CTA.
Note that although the curves for $\theta_{\rm zen}=20$ and $\theta_{\rm zen}=50$
can differ significantly, in particular for the afterglow detections,
the true result should lie in between them as discussed above.
One also sees that 90~\% of the prompt detection bursts have redshifts $z<$2.7--3.4.

\begin{figure}[!htb]
\centerline{\includegraphics[width=0.9\columnwidth]{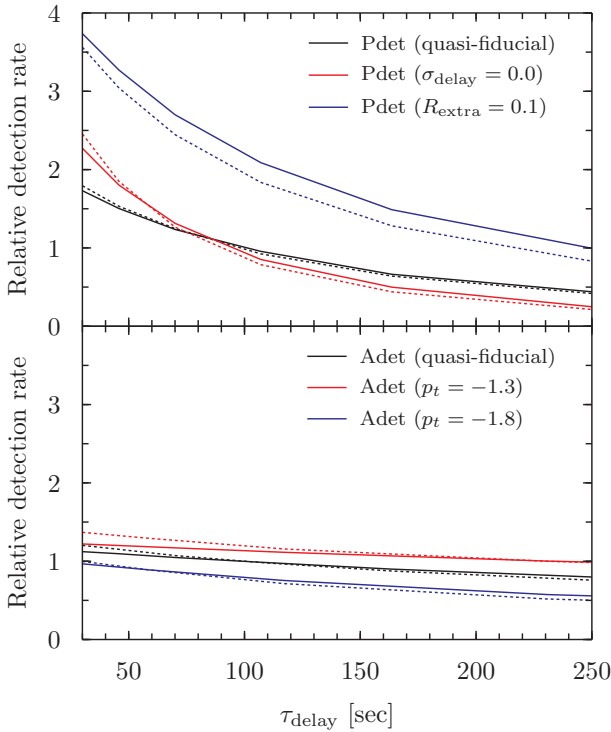}}
\caption{
Detection rate relative to the fiducial result as a function of $\tau_{\rm delay}$.
Solid and dashed curves correspond to $\theta_{\rm zen}=20$~deg and $\theta_{\rm zen}=50$~deg, respectively.
Top panel:
Prompt detection events {\it Pdet}.
Black curves labelled ``quasi-fiducial'' are the case with fiducial parameters except for the mean delay time $\tau_{\rm delay}$.
Red curves labelled ``$\sigma_{\rm delay}=0.0$'' are the case neglecting the dispersion of $T_{\rm delay}$,
and blue curves labeled ``$R_{extra}=0.1$'' are the case with an extra spectral component.
Bottom panel:
Afterglow detection events {\it Adet}.
Black curves labelled ``quasi-fiducial'' are the case with fiducial parameters except for the mean delay time $\tau_{\rm delay}$.
Red and blue curves are the cases with temporal decay index $p_t=-1.3$ and $p_t=-1.8$, respectively.
}
\label{fig:delay-vs-PEAG}
\end{figure}

Fig.~\ref{fig:delay-vs-PEAG}
summarizes the dependence of the detection rate
on $\tau_{\rm delay}$ and other parameters.
The prompt detection rate is quite sensitive to $\tau_{\rm delay}$,
as the relatively steep tail of the duration distribution
implies that the number of bursts satisfying $T_{\rm delay}<T_{90}$ varies strongly.
The addition of an extra spectral component with $R_{\rm extra}=0.1$
increases the detection rate by a factor of $\sim 2$,
regardless of $\tau_{\rm delay}$ or $\theta_{\rm zen}$.
On the hand, the detection rate for afterglows does not depend greatly on $\tau_{\rm delay}$
on account of their long-lasting nature.
Except for some cases with $R_{\rm extra} = 0.1$, $p_E = -1.5$ or $-2.5$,
the detection rates do not differ by more than 20\% for variations in our parameters
(see \cite{Kakuwa12} for details).

We also estimate the detection rate for the case of alerts from SVOM in a simplified way as follows.
Its ECLAIRS instrument will provide alerts with localization $<10^\prime$ at a rate $\sim$ 80~yr$^{-1}$ \citep{Paul11}.
As its energy band (4--250~keV) is similar to {\it Swift} BAT (15--150~keV),
we assume that the duration distribution will be the same,
as presented at the {\it Swift} website \footnote{http://swift.gsfc.nasa.gov/docs/swift/swiftsc.html}.
Since  $\sim 90$~\% of BAT bursts are long GRBs,
and 80~\% of SVOM bursts are expected at $z<6$~\citep{Paul11},
the fraction of SVOM bursts with $T_{90}>2$~sec and $z<5$ is taken to be 70~\%.
For brevity we set $T_{\rm delay}=80$~sec for all SVOM alerts, anticipated to be faster than GBM.
Thus the fraction of long GRBs with $T_{90}>T_{\rm delay}$ is estimated to be $\simeq33$~\%.

Compared to GBM, alerts from SVOM will lead to a larger fraction of bursts that are fainter for CTA,
since it 1) can achieve good localization at lower fluences and 2) is more sensitive to softer bursts, 
including those at higher redshifts that are more affected by EBL attenuation. 
Here we choose not to account for these effects in detail,
but simply assume that the CTA detection efficiency, i. e., the ratio of detectable events ({\it Pdet} or {\it Adet})
to events that are followed up by CTA ({\it CTAobs}), is one-half of that for GBM alerts.
We believe this to be a reasonable approximation; for reference,
if the trigger threshold in peak photon flux for {\it Swift} BAT alerts is set to 0.4~ph~s$^{-1}$cm$^{-2}$ in the 15--150~keV band
(satisfied by 90~\% of actual BAT bursts),
the CTA detection efficiency with $\tau_{\rm delay}=80$~sec
would be about 0.4 times that for GBM alerts with $\tau_{\rm delay}=100$~sec.
In addition, we incorporate a factor of 1.4 enhancement due to anti-solar bias, following \cite{GilmoreGRB}. 

With these assumptions, Table~\ref{table:GRBrate_theory1} shows 
the resulting CTA detection rate for SVOM alerts,
together with {\it Init}, {\it CTAobs} and {\it Pobs} defined analogously to the GBM case.
Compared to GBM alerts, the detection rates for both prompt and afterglow emission are appreciably higher,
even though the frequency of SVOM alerts suitable for CTA follow-up is comparable.
This underscores the importance of GRB facilities with good localizations in the CTA era.


\section{Following up alerts and wide-field mode observations}
\label{sec:followup}

As discussed above, CTA has major scientific potential to advance our understanding of GRBs.
It can both follow up GRBs found with other facilities (Sections \ref{sec:rates},\ref{sec:alert})
and also find GRBs using both standard and survey (i.e. wide-field) modes of operation (Section \ref{sec:widefield}).

\subsection{GRB alerts from satellites and other facilities}
\label{sec:alert}

The study of GRBs depends crucially on localising them in an efficient manner
with a positional accuracy good enough to enable multi-wavelength follow-up.
Launched in 1991, the BATSE instrument on {\it CGRO}
discovered thousands of these explosive events~\cite{Paciesas99}
but with relatively poor localisation accuracy, of order degrees.
The sky and flux distributions of the {\it BATSE} GRBs strongly suggested an extragalactic origin.
Proof, however, only came in in 1997 when the {\it BeppoSAX} satellite~\cite{Boella97},
which could be slew in a few hours, began to detect X-ray afterglows~\cite{Costa97}.
The {\it BeppoSAX} X-ray localisation accuracies (arcmin) were sufficient to search
for optical afterglows~\cite{vanParadijs97}
and soon led to determination of the first GRB redshifts~\cite{Metzger97}.

More recently, the study of GRBs has been revolutionised by the GRB-dedicated {\it Swift} satellite~\cite{Gehrels04}.
Launched in 2004, {\it Swift} can slew very rapidly once it detects a GRB
and use its on-board multiwavelength capability to probe the physics of both the prompt and afterglow emission.
The GRB detection capabilities of {\it Swift} compared to other current and planned satellite facilities
are summarised in Table~\ref{tab:grbsatellites}.
Although {\it Swift} is not the most prolific GRB finder,
it currently provides the most accurate (arcsec), rapid (few minute) localisations.
Thus the number of GRBs for which redshifts and host galaxies have been identified has dramatically increased.
{\it Swift} is also used, via a Target of Opportunity upload, to provide accurate localisations
for GRBs first discovered by other satellites.
The French-Chinese {\it SVOM} satellite \cite{Schanne10,Paul11} will provide similarly accurate
GRB locations during the period when CTA becomes fully operational.

For CTA to be able to respond to an incoming trigger from another facility,
it must be designed to accept such triggers using a standard protocol
such as VOEvent or a GCN notice, and then act accordingly.
The system must be capable of computing visibility constraints
and deciding which CTA telescopes to slew.
It would also be desirable to provide a real-time analysis and information distribution system
that could be used to rapidly (within a minute) communicate to the community
what it has found, again using a standard protocol. 
This capability will also be required for CTA to be used in real-time, transient survey mode
(Section \ref{sec:widefield}).

Some alerts may also be provided by
ground-based air-shower detectors sensitive to TeV gamma-rays such as
HAWC\footnote{http://hawc.umd.edu/} or
LHAASO\footnote{http://english.ihep.cas.cn/ic/ip/LHAASO/}.
For example, HAWC should be able to detect bright bursts similar to GRB 090510 or GRB 090902B
on its own, possibly down to energies as low as 50 GeV \cite{Abeysekara12}.
Despite having less sensitivity and higher energy threshold compared to IACTs,
their much wider FoV and higher duty cycle make them valuable facilities
for discovering VHE transients that may be more or less guaranteed to be detectable
by CTA if followed up sufficiently rapidly.

We also note that some non gamma-ray survey facilities, such as the {\it GAIA}
satellite~\cite{Perryman02}, have the potential to find GRBs via their optical light,
providing additional targets for CTA.

\begin{table*}
\caption{Summary of the main currently operating and near-future planned 
GRB-triggering satellites and their most relevant high energy instruments.
}
\label{tab:grbsatellites}
\begin{center}
\begin{tabular}{l l l l l}
Satellite/instrument & Energy   &  Observed GRB     &  Typical localisation&   Typical \\
               &   range    &  rate (yr$^{-1}$) &  error (radius)  & delay time\\    \hline        
Swift BAT &  15--150 keV &  95    &    few arcmin           &         20 s\\
Swift XRT &  0.3--10 keV      &     90 &       $<$3$''$ &    $\sim$70 s   \\
Fermi GBM &8 keV--40 MeV &    250&     10$^{\circ}$ / 1-3$^{\circ}$ &   20-300 s / 20 min-2 hr \\
Fermi LAT &  20 MeV--300 GeV & 10     &   few deg / 10-60$'$   &     $<$1 min / 4-8 hr  \\     
INTEGRAL IBIS & 20--100 keV & 25--50 &  few arcmin  &               $\sim$60 s\\
Wind-KONUS  &50--200 keV &   100 &              -                      &  1-25 hr\\
Suzaku WAM  & 50--5000 keV & 95   &             -    &                    several hr\\
MAXI-ISS&        2--10 keV       &   $<$10    &  1$^{\circ}$    &                 20 min - few hr\\
IPN      &              various   &     few& arcmin-degrees & 1-1.5 days\\
SuperAGILE&     10--40 keV   & few & few arcmin         & 1-3 hrs\\
SVOM Eclairs&   4--250 keV &     80  &            7$'$    &                      10 s\\
SVOM MXT     &  0.3--7 keV  &    50   &           20$''$    &                    $>$5 min\\     
\end{tabular}
\end{center}
\end{table*}

\subsection{Multiwavelength follow-up}
\label{sec:multiwave}

As GRBs emit at all frequencies their study benefits from a suite of facilities in space and on the ground to follow them up.
Examples of follow-up facilities currently in use or planned in the near future are given in Table~\ref{tab:grbfollowup}.
These facilities can monitor GRBs on timescales ranging from seconds to months or even years after the event.
Fast-response smaller facilities are generally used as filters to identify the most important candidates,
such as potential high-redshift or heavily reddened objects,
which can then be investigated using the larger facilities or space-based assets. 

It would be a valuable addition to the CTA observatory to have an
on-site robotic telescope observing in the optical/near-infrared.
This telescope --- probably of 1--2m class --- could provide real-time monitoring of all CTA science targets
as well as providing rapid follow-up capability for transients such as GRBs.
It will be especially crucial for obtaining redshifts for the bursts,
either on its own or by providing further alerts with good localization to larger telescopes.
The failure to measure a redshift can seriously limit the physics that can be explored
even when a CTA detection is achieved.

In addition to photons, GRBs are prime candidates for the study of
cosmic non-electromagnetic signals that will be studied by
neutrino telescopes such as IceCube~\cite{HK10} (see Section 4.2),
as well as gravitational wave experiments such as Advanced LIGO~\cite{Harry10}.
Gravitational waves may potentially be emitted by GRB progenitors
and correlated with electromagnetic radiation in various ways
(e.g. \cite{Kochanek93, Kobayashi03, Romero10, Fryer11}).
The future simultaneous detection of photons, neutrinos and gravity waves
from a GRB would be one of the greatest achievements in astrophysics.

\begin{table*}
\caption{Summary of the main currently operating and near-future planned GRB follow-up facilities
in alphabetical order, grouped by size and wavelength.
Acronyms are expanded only for the $>$2m optical/near-infrared (nIR) telescopes.
}
\label{tab:grbfollowup}
\begin{center}
\begin{tabular}{llll|llll}                            
6-10m &\multicolumn{3}{l}{Bolshoi Teleskop Altazimuth} & $>$2-5m&\multicolumn{3}{l}{Anglo-Australian Telescope} \\
Optical/nIR&\multicolumn{3}{l}{Gemini (North+South)}   &Optical/nIR&\multicolumn{3}{l}{Bok}                \\
&\multicolumn{3}{l}{Gran Telescopio Canarias}                 &&\multicolumn{3}{l}{Calar Alto}                     \\
&\multicolumn{3}{l}{Hobby-Eberly Telescope}                   &&\multicolumn{3}{l}{Canada-France-Hawaii Telescope}  \\
&\multicolumn{3}{l}{Keck (1-2)}                               &&\multicolumn{3}{l}{ESO 3.6m and 2.2m}              \\
&\multicolumn{3}{l}{Large Binocular Telescope}                &&\multicolumn{3}{l}{Hale Telescope}                 \\
&\multicolumn{3}{l}{Magellan (Baade+Clay)}                    &&\multicolumn{3}{l}{Lick Shane Telescope}                \\
&\multicolumn{3}{l}{South African Astronomical Observatory}          &&\multicolumn{3}{l}{Magnum Mirror Telescope}        \\
&\multicolumn{3}{l}{Subaru}                                   &&\multicolumn{3}{l}{New Technology Telescope}       \\
&\multicolumn{3}{l}{Very Large Telescope (1-4)}               &&\multicolumn{3}{l}{Nordic Optical Telescope}       \\
&&&                                                           &&\multicolumn{3}{l}{Telescopio Nazionale Galileo}      \\
&&&                                                           &&\multicolumn{3}{l}{UK InfraRed Telescope}            \\
&&&                                                           &&\multicolumn{3}{l}{William Herschel Telescope}            \\
&&&                                                           &&\multicolumn{3}{l}{WIYN Telescope}                          \\ \hline
Up to 2m&   ABT                 & LOAO         &RAPTOR          &Sub-mm/&ALMA  &Space& Astro-H \\
Optical/nIR&Aristarchos         &Lulin         &REM             &mm/radio&APEX &~-based&Astrosat \\            
&BOOTES                         &Maidanak      &ROTSE           &&ASKAP        &&Chandra \\           
&CrAO                           &MASTER        &ROVOR           &&CARMA   &&GEMS \\               
&Danish                         &McDonald      &RTT             &&GBT     &&Herschel \\         
&Faulkes                        &MITSuME       &SARA            &&JCMT    &&HST \\              
&FLWO                           &MOA           &SkyMapper       &&LOFAR   &&NuSTAR \\           
&GRAS                           &Mondy         &SMARTS          &&MeerKAT &&SRG-eROSITA \\      
&GRT                            &Newcastle     &SuperLOTIS      &&e-MERLIN&&SVOM-VT \\           
&INT                            &OSN           &Tarot           &&Mullard &&Swift-UVOT  \\        
&KAIT                           &PAIRITEL      &THO             &&PdB     &&UFFO\\              
&Kanata                        &Palomar 60$''$&TLS             &&SMA     &&XMM-Newton \\ 
&Konkoly                        &Pi of the Sky &Xinglong TNT    &&EVLA    && \\
&Lightbuckets                 &PROMPT        &Yunnan          &&WSRT    && \\    
&Liverpool                      
\end{tabular}
\end{center}
\end{table*}

\subsection{Observations of GRBs and other transients in standard or wide-field modes}
\label{sec:widefield}

CTA also has the potential to open up a unique discovery space for GRBs at very high energies
by finding GRBs in its own right, including those with unusual spectral energy distributions
which did not trigger gamma-ray instruments.
GRBs discovered by CTA itself would lead to complete coverage of the prompt emission phase,
addressing important questions on the origins of both long and short-duration GRBs and on Lorentz invariance.
Such observations are in fact the only way for CTA to
detect or set limits on very high energy prompt emission from
short GRBs, which, in the era of gravitational wave capability with
Advanced LIGO, will be of great importance.
Note that operation in such observing modes requires a real-time analysis routine which identifies new sources,
reports new transients to the community upon detection as described in Section \ref{sec:alert}
and may override the current observation.

{\bf Standard observing mode.}
While observing a science target CTA can also be searching for transients within the field of view.
Although each field of view is modest in standard mode, the total integration time would be large,
CTA would be operating at maximum sensitivity and can be used during the entire duty-cycle.
 
{\bf Wide-field mode.}
In wide-field or survey mode the telescopes are offset and spread over a wide field of view to maximise solid angle \cite{Dubus12}.
Although GRBs occur equally in all regions of the sky,
those around the Galactic plane are affected by interstellar absorption
that hampers X-ray and optical followup efforts, necessary for good localization and redshift determination.
Thus spreading the MSTs centered at a high Galactic latitude may be the most promising strategy.
The much increased solid angle implies a greater GRB rate at the cost of a decrease in sensitivity,
so would be complementary to transient searches in the standard observing mode.
While a substantial fraction of time would have to be spent in a wide-field mode,
this need not be done consecutively and could be done in parallel
while the LSTs/SSTs are observing other sources, providing a low-cost transient survey mode for CTA.
This mode also permits a large-area sky survey to be built up over time.

We can estimate the rates for GRBs that are potentially observable in wide-field mode,
based on the sky-rates derived from current GRB-triggering satellites
{\it Swift} and {\it Fermi} (Table \ref{tab:grbsatellites}).
In contrast to the case of responding to external triggers (Section \ref{sec:rates}),
we would not need to consider satellite time delays for reporting a GRB, nor the size of localisation errors.
We assume a configuration of 25 MSTs which cover 2.5\% of the sky
(minimal overlap with no gaps in the field of view) and have a duty cycle of 10\%.
Multiplying the observed GRB rate to account for the instrument solid angle
and the fraction of time spent able to trigger on GRBs,
we arrive at all-sky annual GRB rates of $\sim 800$ and $\sim 600$ for BAT and GBM respectively.
The rate of {\it Swift}- and {\it Fermi}-like GRBs
that are observable by CTA in wide-field mode is roughly 2--3 per year,
or 0.2--0.3 events per 100 hours of observation.
Of this, the fraction that is actually detectable requires knowledge of the sensitivity in this mode
near the energy threshold on short exposure times
that is currently being studied by the MC group and will be available in the future.

A wide-field survey will be of great interest to the wider scientific community,
encompassing the study of all transient phenomena,
e.g. supernovae and accretion powered sources including active galaxies and X-ray binaries.
Further value can be added if the timing and/or field of view were to coincide with comparable surveys
at other wavelengths providing a means of characterising newly discovered phenomena.
(For other aspects of surveys with CTA, see \cite{Dubus12}.)


\section{Conclusions and outlook}
\label{sec:conc}

With high photon statistics measurements of their multi-GeV spectra and temporal variability,
the science cases that can be explored by observing GRBs with CTA are varied and far-reaching.
In addition to the many mysteries surrounding the physics of GRBs themselves, they include
the origin of ultra-high-energy cosmic rays and prospects for high-energy neutrinos via hadronic gamma-ray signatures,
the cosmic history of star formation, black hole accretion and intergalactic reionization
via extragalactic background light attenuation features, and
precision tests of special relativity via searches for energy-dependent delays, etc.
Further simulations of spectra and light curve observations 
in greater detail than presented here will be helpful to better quantify the prospects.
In view of the modest expected detection rate of a few bursts per year,
a key issue will be to assure that all GRB alerts observable with CTA
are followed up sufficiently rapidly under stable conditions.
Achieving an energy threshold as low as possible will also be crucial,
not only for enhancing the detection rate,
but also to improve the photon statistics per detection,
as well as to attain the broadest energy range over which
time-resolved spectra or energy-dependent light curves can be studied
to address the science cases discussed here.
The principal GRB alert facilities in the CTA era are likely to be SVOM and {\it Fermi} GBM,
especially if the latter's localization accuracy can be appreciably improved by that time,
and possibly also Swift if its instrumental performance and funding can be maintained.
Other potential satellites such as the All-Sky Transient Astrophysics Reporter (A-STAR)
will undoubtedly be of great value.
An onsite optical-infrared telescope would be valuable for ensuring
that redshift measurements are performed for all CTA bursts.
Dedicated, wide-field survey mode observations
will be a unique way to discover GRBs with CTA alone including short GRBs and to detect long GRBs from their onset,
as well as to conduct an unbiased search for transients.

\section*{Acknowledgements}
We are grateful to Masanori Ohno for providing us with {\it Fermi} LAT data on the light curve of GRB 080916C.
We thank comments from Felix Aharonian and Elisabetta Bissaldi.
This work is supported in part by Grants-in-Aid from the
Ministry of Education, Culture, Sports, Science and Technology (MEXT) of Japan,
Nos. 22540278 and 24340048 (SI), Nos. 24103006, 24000004, 22244030, 22244019, 21684014 (KI),
and No. 21740184 (RY).
YI, KM and KT are supported by JSPS Research Fellowships for Young Scientists.
JPO acknowledges financial support from the UK Space Agency.
RS is supported by a Royal Society fellowship.
DW is supported by ? 262053 and from US NSF (PHY-0970134).
The research leading to these results has received funding from the
European Union's Seventh Framework Programme ([FP7/2007-2013]
[FP7/2007-2011]) under grant agreement no 262053.

\end{document}